\documentclass[11pt]{article}
\usepackage{jheppub}

\usepackage[space]{grffile}
\usepackage{pdfpages}

\usepackage{slashed,physics}
\usepackage{graphicx}
\usepackage{amsmath,amssymb,graphicx}
\usepackage{textcomp} 
\usepackage{gensymb} 
\usepackage{epsf,color}
\usepackage[dvipsnames]{xcolor}
\usepackage{hyperref}
\usepackage{cancel}
\usepackage{framed}
\usepackage{hyperref}
\usepackage{float}
\usepackage{multirow}
\usepackage{subfigure}
\definecolor{nicered}{rgb}{0.5,0.,0.}
\definecolor{nicegreen}{rgb}{0.,0.5,0.}
\definecolor{niceblue}{rgb}{0.,0.,0.5}
\hypersetup{colorlinks,citecolor=nicegreen,linkcolor=nicered,urlcolor=niceblue}
\numberwithin{equation}{section}
\newcommand{\beq}{\begin{equation}}
\newcommand{\eeq}{\end{equation}}
\newcommand{\bea}{\begin{eqnarray}}
\newcommand{\eea}{\end{eqnarray}}
\newcommand{\bear}{\begin{eqnarray}}
\newcommand{\eear}{\end{eqnarray}}

\newcommand{\TeV}{\text{ TeV}}

\newcommand{\mm}{\mu^+\mu^-}
\newcommand{\nn}{\nu_{\mu}\bar{\nu}_{\mu}}

\newcommand{\kmu}{\kappa_{\mu}}

\newcommand{\calS}{\mathcal{S}}

\newcommand{\ba}{\begin{array}}
\newcommand{\ea}{\end{array}}

\title {Precision Test of the Muon-Higgs Coupling at a High-energy Muon Collider}

\author[a]{Tao Han,}
\affiliation[a]{Pittsburgh Particle Physics, Astrophysics, and
  Cosmology Center, Department of Physics and Astronomy, University of
  Pittsburgh, Pittsburgh, PA 15206, USA} 
\author[b]{Wolfgang Kilian,}
\affiliation[b]{Department of Physics, University of Siegen,
  Walter-Flex-Stra{\ss}e 3, 57068 Siegen, Germany}
\author[b]{Nils Kreher,}
\author[a]{Yang Ma,}
\author[c]{J\"urgen Reuter,}
\affiliation[c]{Deutsches Elektronen-Synchrotron DESY, Notkestra{\ss}e 85, 22607 Hamburg,
Germany}
\author[b]{Tobias Striegl,}
\author[a]{and Keping Xie}

\emailAdd{than@pitt.edu}
\emailAdd{kilian@physik.uni-siegen.de}
\emailAdd{nils.kreher@uni-siegen.de}
\emailAdd{mayangluon@pitt.edu}
\emailAdd{juergen.reuter@desy.de}
\emailAdd{tobias.striegl@physik.uni-siegen.de}
\emailAdd{xiekeping@pitt.edu}

\preprint{DESY 21-086, PITT-PACC-2110}

\abstract{
   We explore the sensitivity of directly testing the muon-Higgs  coupling at a high-energy muon collider. This is strongly motivated if there exists new physics that is not aligned with the Standard Model Yukawa interactions which are  responsible for the fermion mass generation. We illustrate a few such examples for physics beyond the Standard Model. With the accidentally small value of the muon Yukawa coupling and its subtle role in the high-energy production of multiple (vector and Higgs) bosons,  we  show that it is possible to measure the muon-Higgs coupling to an  accuracy of ten percent for a 10 TeV muon collider and a few percent for a 30 TeV machine by utilizing the three boson production, potentially sensitive to a new physics scale about $\Lambda \sim 30-100$ TeV.
}

\begin{document}
\maketitle

\section{Introduction}
\label{sec:intro}

The Standard Model (SM) of elementary particle physics is constructed
based on a non-Abelian gauge theory of SU(3)$_{\rm C} \otimes$
SU(2)$_{\rm L}\otimes$U(1)$_{\rm Y}$, that has been experimentally
verified with a high accuracy to the highest energies accessible to
date \cite{Zyla:2020zbs}. On the other hand, there is mounting
evidence from observations for the need of new physics beyond the SM,
such as the dark matter, neutrino mass generation, and the
matter/antimatter asymmetry. 

Unlike the past decades, at the moment we are lacking well-defined
traces of where to look for new physics. While there are many loose ends in the SM of particle physics and cosmology, however, there is no clear indication at what energy scales new phenomena would appear below the Planck scale. This gives us the task to use all available tools to search for new phenomena,
particularly all the discovered particles as vehicles for our
searches. Especially, the scalar boson discovered in
2012~\cite{Aad:2012tfa,Chatrchyan:2012xdj} which closely resembles the SM Higgs boson is very well suited for beyond the Standard Model (BSM)
searches~\cite{deFlorian:2016spz}. 
Currently, the couplings of the Higgs boson to the third generation SM fermions have been established with a precision of $10\% -20\%$ (for an overview of the current status and projections, see {\it e.g.}~\cite{deBlas:2019rxi}).
The high-luminosity phase of the
LHC will study the properties of this particle and its couplings to a precision at a few percent level \cite{ATL-PHYS-PUB-2018-054,CMS-PAS-FTR-18-011}.
The next collider facility will most likely be a Higgs factory~\cite{EuropeanStrategyforParticlePhysicsPreparatoryGroup:2019qin,EuropeanStrategyGroup:2020pow} in the form of an electron-positron
collider running at or slightly above the $ZH$
threshold, such as the International Linear Collider (ILC) \cite{Baer:2013cma,
Behnke:2013lya}, the Future Circular Collider (FCC-ee) \cite{Abada:2019zxq}, the Circular Electron-Positron Collider (CEPC)  \cite{CEPCStudyGroup:2018ghi}, or the Compact Linear Collider (CLIC) at higher energies \cite{Aicheler:2012bya,CLIC:2016zwp} to achieve a per-mille level accuracy for the Higgs couplings to $W^+W^-,ZZ,\gamma\gamma,gg$ and $b\bar b, \tau\bar\tau, c\bar c$, as well as the invisible decay mode.

However, there will still be parts of the Higgs sector left unexplored or measured with low precision because it can only be probed with very rare processes for which there are too low rates at a Higgs factory and the LHC measurements (or searches) suffer from large systematic uncertainties due to the challenging experimental environment. To this class belong the couplings to the first and second generations of fermions.
The Higgs mechanism in the SM provides the mass for all elementary particles, and thus specifies the form of their interactions associated with the electroweak symmetry breaking (EWSB).  With only a single SU(2)$_L$ Higgs doublet and the minimal set of interactions at the renormalizable level, the Yukawa couplings of SM fermions are proportional to the respective particle masses, and thus exhibit a large hierarchy.
It would be desirable to achieve a better 
precision for the measurement of the Yukawa couplings of the light fermions, since this would be a direct and important test whether the Higgs mechanism as implemented in the SM provides the masses for all SM fermions, or whether it is a mixture of two (or more) mechanisms. 
Because of the small Yukawa couplings for light fermions predicted in the SM, any small deviation due to BSM physics may result in a relatively large modification to those couplings.

The next target is the Higgs-muon coupling. The recent evidence for the
$H\to\mu^+\mu^-$ decay at ATLAS and CMS indicates that the Yukawa
coupling is present within the predicted order of magnitude  \cite{Sirunyan:2020two,Aad:2020xfq}.  However, the results are not yet at the $5\sigma$ level for discovery, and thus leaves room for $O(100\%)$ corrections. Also, the measurement is insensitive to the sign of the coupling. 
According to the current experimental
projections, by the end of the high-luminosity runs of the LHC in the late 2030s the muon Yukawa coupling could be measured with an accuracy of about several tens of percent 
\cite{ATL-PHYS-PUB-2014-016} in a model-dependent way. 
This situation might not be improved very much neither at the Higgs factory due to the limited rate, nor at a high-energy hadron collider like the
FCC-hh~\cite{Abada:2019lih,Benedikt:2018csr}, due to the systematics and the model-dependence. 
Thanks to the technological development~\cite{Delahaye:2019omf}, a renewed idea that has recently gathered much momentum
is the option of a high-energy muon collider that could reach the multi-(tens of) TeV regime with very high 
luminosity \cite{Bartosik:2020xwr,Schulte:2020xvf,Long:2021upy}. %
It has been demonstrated in the recent literature that a high-energy muon collider has great potential for new physics searches at the energy frontier from direct $\mu^+\mu^-$ annihilation and a broad reach for new physics from the rich partonic channels \cite{Han:2020uid,Costantini:2020stv,Buttazzo:2020uzc,Han:2021kes,Buarque:2021dji}, as well as precision measurements for SM physics \cite{Han:2020pif} and beyond \cite{Han:2020uak,Han:2021udl,Capdevilla:2020qel,Yin:2020afe,Capdevilla:2021rwo,Liu:2021jyc,Gu:2020ldn,Huang:2021nkl,Capdevilla:2021fmj}. Of particular importance is the connection between the muon collider expectation and the tantalizing hint for new physics from the muon $g-2$ measurement  \cite{Muong-2:2006rrc,Muong-2:2021ojo}.

In this paper, we propose one unique measurement and BSM search in the Higgs sector which serves as a paradigm example for exploiting a high-energy muon collider, namely the direct  measurement of the muon Yukawa coupling. At a high-energy $\mm$ collider, one probes the coupling at a much higher energy scale and it may reach some sensitivity to new physics with scale-dependent effects.
Unlike the precision measurements at low energies where one probes the virtual quantum effects, our proposal is to directly measure the muon coupling associated with its mass generation. Our search strategy is generally applicable to other new physics
searches involving final states of charged leptons and jets, that may
provide general guidance for future considerations.  

The rest of the paper is organized as follows. We first present a brief overview and motivation for the importance of
studies of the muon Yukawa coupling in Sec.~\ref{sec:setup}. 
In Sec.~\ref{sec:muonhiggs}, we examine the renormalization group (RG)-induced scale dependence of the couplings. This is important to relate a measured quantity in a high-energy collider setup to the low-scale value.
In Sec.~\ref{sec:MuY}, we construct an effective field theory (EFT) setting to
discuss possible deviations of the muon Yukawa coupling from its SM
value. We present a few paradigm examples of modifications of the muon-Higgs coupling from its SM Yukawa value. 
In Sec.~\ref{sec:smeft} we then discuss different EFT parameterizations, constraints from unitarity limits in Sec.~\ref{sec:unitarity}, and consequences for ratios of different production cross sections in Sec.~\ref{sec:ratios}. 
It sets the theoretical frame for our phenomenological studies in Sec.~\ref{sec:Pheno}, where 
we analyze the collider sensitivity for the determination of the muon Yukawa coupling at a high energy muon collider, before we conclude in Sec.~\ref{sec:summary}.

\section{Theoretical Considerations for the Muon Yukawa Coupling}
\label{sec:setup}

\subsection{Illustrations of the running of the Muon Yukawa Coupling}
\label{sec:muonhiggs}

When testing the muon-Higgs Yukawa coupling, it is necessary to properly take into account the energy-scale dependence of the coupling, which is a fundamental prediction in quantum field theory. The specific form of this running depends on the particle spectrum and their interactions in the underlying theory. In the electroweak sector of the SM, the dominant contribution to the renormalization group (RG) running is the top Yukawa coupling, followed by the strong and EW gauge interactions.

\begin{figure}[htb]
\centering
\includegraphics[width=0.8\textwidth]{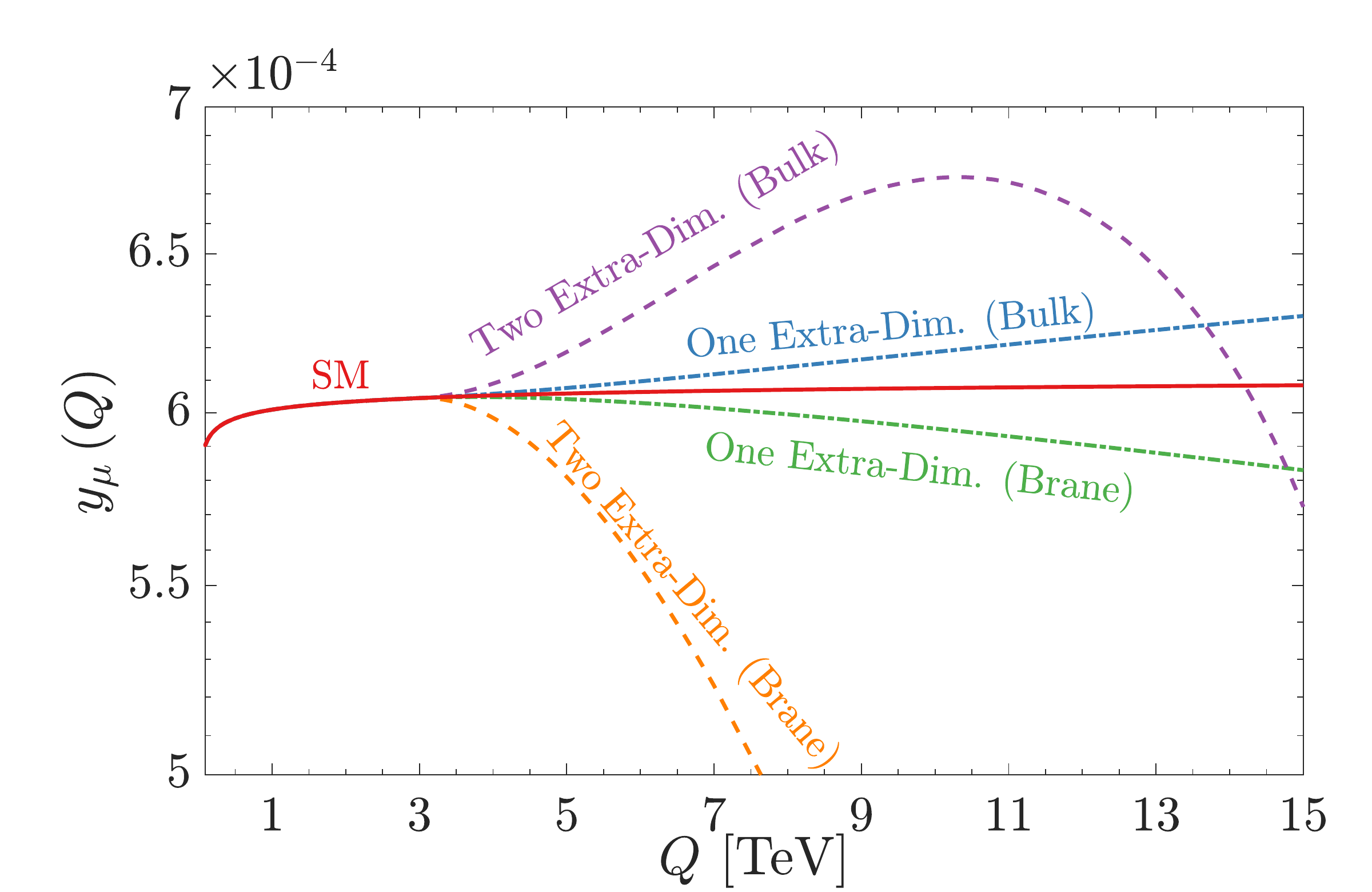}
\caption{LO RGE running of the muon Yukawa $y_\mu$ coupling as a function of the energy scale $Q$, in the SM (red solid). In the extra-dimensional scenarios (with inverse radius $1/R = 3$ TeV), we consider 1) Bulk: all fields propagating in the bulk, and 2) Brane: all matter fields localized to the brane. 
}
\label{fig:ym_run}
\end{figure}

\begin{figure}[htb]
  \centering
  \includegraphics[width=0.8\textwidth]{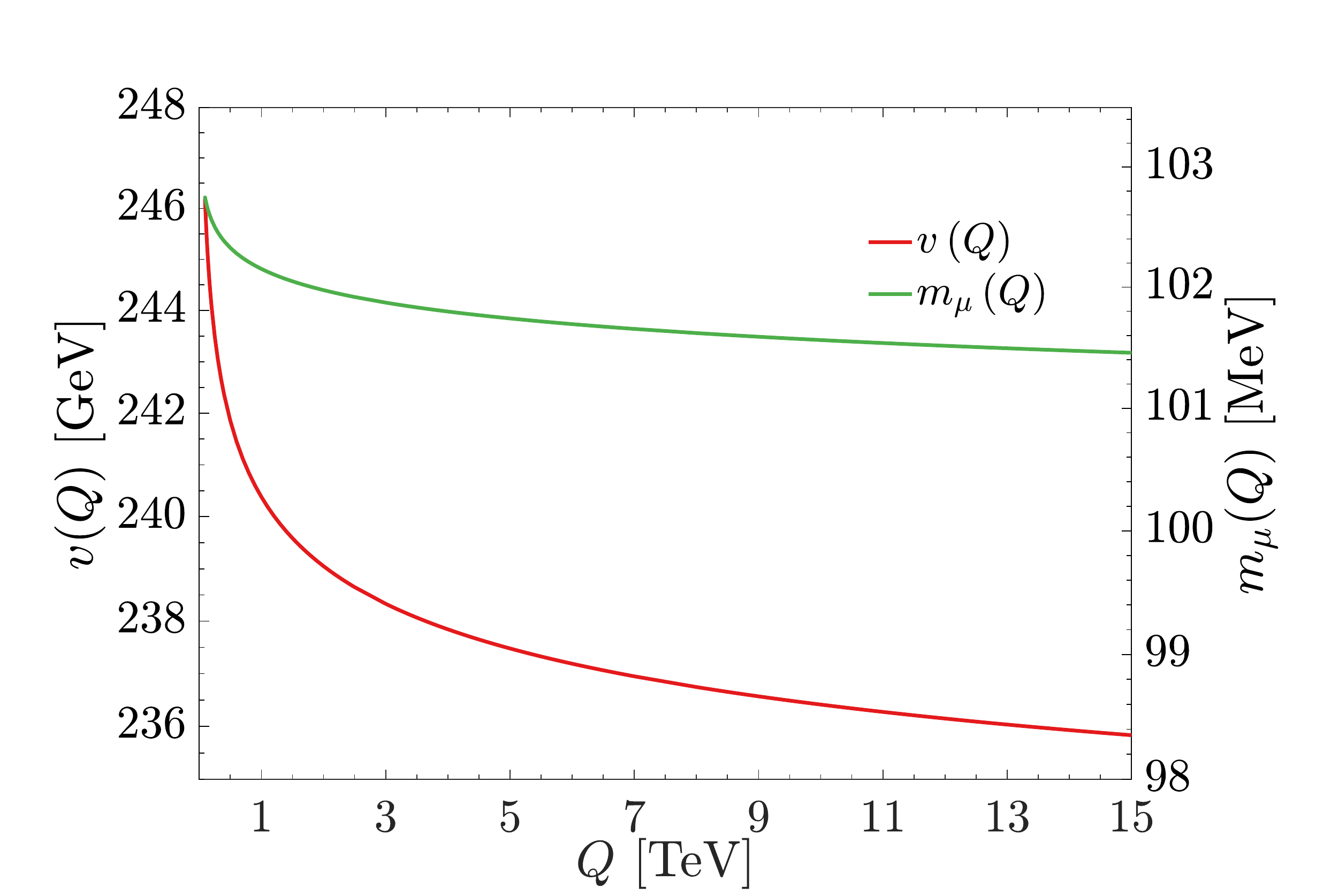}
  \caption{LO RGE running of SM vacuum expectation value $v$ (left scale) and  muon mass $m_\mu$ (right scale) as functions of the energy scale $Q$.
  }
  \label{fig:vev_run}
\end{figure}

For the sake of illustration, the coupled renormalization group equations (RGEs) of Yukawa couplings $y_\mu, \, y_t$, vacuum expectation value $v$, and gauge couplings $g_i$ are given in the $\overline{\rm MS}$ scheme at leading order (LO) in one-loop by \cite{Machacek:1983tz,Machacek:1983fi,Arason:1991hu,Arason:1991ic,Arason:1992eb,Castano:1993ri,Grzadkowski:1987tf} 
\begin{eqnarray}
\beta_{y_t}&=& \frac{\dd y_t}{\dd t} = \frac{y_t}{16 \pi^2} \left (\frac{9}{2}y_t^2 - 8 g_3^2 - \frac{9}{4} g_2^2 - \frac{17}{20} g_1^2 \right), \\
\beta_{y_\mu}&=& \frac{\dd y_\mu}{\dd t} = \frac{y_\mu}{16 \pi^2} \left (3y_t^2 - \frac{9}{4}(g_2^2 + g_1^2) \right), \\
\beta_{v}&=& \frac{\dd v}{\dd t} = \frac{v}{16 \pi^2} \left(\frac{9}{4} g_2^2+\frac{9}{20} g_1^2-3 y_t^2 \right), \\
\beta_{g_i} &=& \frac{\dd g_i}{\dd t} = \frac{b_i g_i^3}{16 \pi^2},
\end{eqnarray}
with $t=\ln(Q/M_Z)$ and the coefficients  $b_i$ for the gauge couplings $(g_1,g_2,g_3)$ given as
\begin{align}
b_i^{\rm SM} = & (41/10,-19/6,-7) .
\end{align}
We show the LO RGE running of the muon Yukawa $y_\mu$ in the SM in Fig.~\ref{fig:ym_run} (red solid curve) and the SM vacuum expectation value $v$ in Fig.~\ref{fig:vev_run} (left axis) as functions of the energy scale $Q$, respectively. With the relation 
$$m_\mu(Q)=y_\mu(Q) v(Q)/\sqrt 2,$$ 
we also show the running of the muon mass, $m_\mu(Q)$, in Fig.~\ref{fig:vev_run} (right axis). 
At the energy scales accessible in near future colliders, the change in $y_\mu$ is observed to be rather small, for example,  $y_\mu(Q=15 {~\rm TeV})$ is found to be around $3\%$ smaller compared to $y_\mu(M_Z)$. Similarly, $v\ (m_\mu)$ runs down by about $4\%$ ($2\%$).

New states appearing in beyond SM scenarios can modify the running of the relevant gauge and Yukawa couplings. Generically, the beta function for a coupling $\lambda$ is given as
\begin{equation}
\beta_\lambda = \beta_\lambda^{\rm SM} + \sum_{\rm s: ~massive ~new ~states} \theta(Q-M_s) \, \times\, N_s \beta_{s,\lambda}^{\rm NP}\;,
\end{equation}
where $\beta_\lambda^{\rm SM}$ is the SM beta function, and $\beta_{s,\lambda}^{\rm NP}$ represents the contribution of a new heavy state $s$ of mass $M_s$, with $N_s$ number of degenerate degrees of freedom. The theta function encodes the fact that the effect of new heavy states is included in the RG running once the energy scale $Q$ is above the threshold $M_s$, ignoring here for simplicity the effect of threshold corrections.

In extensions of the SM, the muon-Higgs Yukawa coupling could also be affected both at the tree level and at the quantum level. In addition, the Higgs sector may show a rich flavor structure. In flavor-sensitive Higgs models, the SM prediction for the Yukawa
couplings is lost, and the Yukawa couplings become free model
parameters.  The physical coupling of the SM Higgs to muons may be 
larger or smaller than its expected SM value.  In principle, it could be
completely absent, such that the muon mass is generated by other means.
The assumption we make for the study in this paper is that the muon
Yukawa coupling is a free parameter, as the mass generation for the
muon is in general a mixture of the SM mechanism and a yet-unknown mechanism. A typical example for this is a Two-Higgs doublet model
(2HDM), or in a general multi-doublet model, that generates
third-generation Yukawa couplings, while the second generation
couplings are from a different sector (a sample implementation of such
a mechanism can be found in~\cite{Altmannshofer:2015esa}). Clearly,
the LHC offers also some opportunities to probe first and second
generation Higgs Yukawa couplings to light
quarks~\cite{Soreq:2016rae}, which applies mostly to the Higgs charm
Yukawa coupling~\cite{Bodwin:2013gca,Kagan:2014ila,Perez:2015aoa,
Bishara:2016jga}, and maybe even strange tagging is possible at a
future Higgs factory~\cite{Duarte-Campderros:2018ouv}.
In weakly-coupled theories, the running effects for the muon-Yukawa coupling are rather moderate, similar in size to that in the SM. We will not show it separately.

An interesting question is also whether there could be considerable CP
violation in the Higgs Yukawa sector beyond CKM, where there are
bounds e.g.~for the electron Yukawa
coupling~\cite{Altmannshofer:2015qra}. Though it is perfectly possible
in our setup in Sec.~\ref{sec:MuY} to discuss CP-violating operators
for the muon Yukawa couplings, such a study is beyond the scope of
this current paper.

We add the remark that additional, flavor-dependent,
higher-dimensional operators that are responsible for a deviation of
the SM muon Yukawa coupling could easily lead to flavor-violating
Yukawa couplings that induced $H \to e\mu$. This has been studied
e.g. in~\cite{Harnik:2012pb}, however, we are not further
investigating such flavor-violating processes in this paper. The EFT
setup for our study is presented in detail in the next section. 

Large modifications to the running couplings compared to the SM case are not expected in four-dimensional quantum field theories essentially due to the logarithmic nature of the running. 
A qualitatively different scenario however is obtained if there is a tower of new physics states modifying the RGEs, asymptotically leading to a power-law running of the Yukawa coupling~\cite{Dienes:1998vh,Dienes:1998vg}. 
This four-dimensional description is equivalent to a theory with compactified flat extra space-like dimensions, with gauge and/or matter fields propagating in the higher-dimensional bulk.
To illustrate this, we consider two scenarios of compactified flat extra-dimensions~\cite{Appelquist:2000nn}: a 5D model with the extra-dimension compactified on an $S_1/Z_2$ orbifold, and a 6D model with the two extra dimensions compactified on a square $T^2/Z_2$ orbifold~\cite{Appelquist:2000nn,Appelquist:2001mj}. In both models, we consider two cases: 1) all SM fields propagating in the bulk and 2) the SM gauge fields to be propagating in the bulk, with the matter fields of the SM restricted to the brane  \cite{Bhattacharyya:2006ym,Cornell:2012qf,Blennow:2011mp,Kakuda:2013kba,Abdalgabar:2013oja}. 
The beta functions of the gauge couplings in such scenarios are given as:
\begin{align}
b_i^{\rm 5D} = & b_i^{\rm SM} + (S(t)-1) \times \left[\left(\frac{1}{10},-\frac{41}{6},-\frac{21}{2}\right)+\frac{8}{3} \eta \right] \nonumber\\
b_i^{\rm 6D} = & b_i^{\rm SM} + (\pi S(t)^2-1) \times \left[\left(\frac{1}{10},-\frac{13}{2},-10\right)+\frac{8}{3} \eta \right].
\end{align}
Here, $S(t)$ counts the number of degrees of freedom $S(t)=e^{t}R$, $R$ being the radius of the extra dimension, $\eta$ being the number of generations of fermions propagating in the bulk. 
The corresponding one-loop RGE equations for the Yukawa couplings $y_t,\, y_\mu$ in the extra-dimensional scenarios are as follows \cite{Cornell:2011fw,Cornell:2012qf,Abdalgabar:2013oja}

\begin{subequations}
\begin{align}
\frac{dy_t}{d t} = & \beta_{y_t}^{\rm SM}  +\frac{y_t}{16\pi^2} 2(S(t)-1)\left(\frac{3}{2} y_t^2 - 8 g_3^2 - \frac{9}{4} g_2^2 - \frac{17}{20} g_1^2\right), &{\rm 5D~Brane}, \\
\frac{dy_\mu}{d t} = & \beta_{y_\mu}^{\rm SM}  -\frac{y_\mu}{16\pi^2} 2(S(t)-1)\left(\frac{9}{4} g_2^2+ \frac{9}{4} g_1^2\right), &{\rm 5D~Brane}, \\
\frac{dy_t}{d t} = & \beta_{y_t}^{\rm SM}  +\frac{y_t}{16\pi^2} (S(t)-1)\left(\frac{15}{2} y_t^2 - \frac{28}{3} g_3^2 - \frac{15}{8} g_2^2 - \frac{101}{120} g_1^2\right), &{\rm 5D~Bulk}, \\
\frac{dy_\mu}{d t} = & \beta_{y_\mu}^{\rm SM}  +\frac{y_\mu}{16\pi^2} (S(t)-1)\left(6y_t^2-\frac{15}{8} g_2^2- \frac{99}{40} g_1^2\right), &{\rm 5D~Bulk}. 
\end{align}
\end{subequations}

\begin{subequations}
\begin{align}
\frac{dy_t}{d t} = & \beta_{y_t}^{\rm SM}  +\frac{y_t}{16\pi^2} 4\pi(S(t)^2-1)\left(\frac{3}{2} y_t^2 - 8 g_3^2 - \frac{9}{4} g_2^2 - \frac{17}{20} g_1^2\right), &{\rm 6D~Brane}, \\
\frac{dy_\mu}{d t} = & \beta_{y_\mu}^{\rm SM}  -\frac{y_\mu}{16\pi^2} 4\pi(S(t)^2-1)\left(\frac{9}{4} g_2^2+ \frac{9}{4} g_1^2\right), &{\rm 6D~Brane}, \\
\frac{dy_t}{d t} = & \beta_{y_t}^{\rm SM}  +\frac{y_t}{16\pi^2} \pi(S(t)^2-1)\left(9 y_t^2 - \frac{32}{3} g_3^2 - \frac{3}{2} g_2^2 - \frac{5}{6} g_1^2\right), &{\rm 6D~Bulk}, \\
\frac{dy_\mu}{d t} = & \beta_{y_\mu}^{\rm SM}  +\frac{y_\mu}{16\pi^2} \pi(S(t)^2-1)\left(6y_t^2-\frac{3}{2} g_2^2- \frac{27}{10} g_1^2\right), &{\rm 6D~Bulk}. 
\end{align}
\end{subequations}
We see from Fig.~\ref{fig:ym_run} that in the presence of such a tower of new states, the running of $y_\mu$ can be substantially altered for both the 5D  (dot-dashed curves), and 6D (dashed curves) models.  
We note that the effects only become significant when close or above the new physics threshold, $1/R\sim 3$ TeV in our illustration. Above the threshold, the other more direct effects from the existence of the extra dimensions may be observable as well and a coordinated search would be beneficial.

We conclude that while in the SM the energy dependence of the $y_\mu$ is a minor effect, there are viable models where the value and the running of this quantity could both follow completely different patterns, as illustrated above with extra-dimensional scenarios. In the next subsection, we will extend this direction in the EFT framework.

\subsection{EFT Description of an Anomalous Muon Yukawa Coupling}
\label{sec:MuY}

In a purely phenomenological ansatz, if small modifications of the SM Lagrangian exist, they should be detectable most easily in interactions which are accidentally suppressed in the SM, and at the same time are unaffected by large radiative corrections.  The muon mass and the associated production and decay
processes perfectly fit this scenario. In this spirit, we introduce  representative new interactions in form of a modification of this muon mass parameter, without referencing a specific model
context.  The modification is supposed to be tiny in absolute terms, but
nevertheless becomes significant if compared with the SM muon Yukawa coupling
which has a numerical value of less than $10^{-3}$.
A few well-motivated physics scenarios with a modification of the SM can be constructed as we will discuss next. They may describe rather different underlying dynamics, but represent physically equivalent calculational frameworks in the perturbative regime.


\subsubsection{The Yukawa interaction in the HEFT parameterization}
\label{sec:heft}

In the Higgs Effective Theory (HEFT)~\cite{Coleman:1969sm,Callan:1969sn,Weinberg:1980wa,Appelquist:1980vg,Longhitano:1980tm,Dobado:1990zh} or non-linear chiral-Lagrangian description, the scalar sector consists of a physical singlet Higgs boson together with unphysical triplet Goldstone bosons associated with the EW symmetry breaking. The latter
isolate the contributions of longitudinally polarized vector bosons.  This property can be formalized as the Goldstone-boson Equivalence Theorem
(GBET)~\cite{Chanowitz:1985hj,Gounaris:1986cr}:
\begin{center}
\raisebox{-.45\height}{
\includegraphics[width=.2\textwidth]{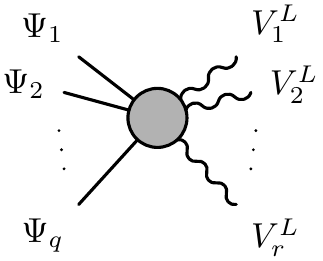}}
=
\raisebox{-.45\height}{
\includegraphics[width=.2\textwidth]{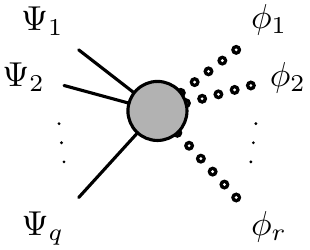}}
$\;+\;\mathcal O \left(\frac{m}{\sqrt s}\right)$
\end{center}
Here, $V^L_k$ denotes a longitudinal EW vector boson, $\phi_k$ the corresponding Goldstone boson, and $\Psi_k$ any possible SM fermion. This denotes that fact that matrix elements for multi-boson final states including vector bosons are dominated in the high-energy limit by their longitudinal component 
\begin{equation} 
  \varepsilon^{\mu}_L(p)=\frac{p^{\mu}}{m}+v_p^{\mu} \quad , 
\end{equation} 
where $v^\mu_p\sim 
\mathcal{O}(m/ \sqrt s)$ is a
four-vector depending on the boson momentum. According to~\cite{Dobado:1997jx} the GBET in an EFT framework takes the form 
\begin{align}
\mathcal M(V^L_1,\dots, V^L_r, \mathbf{\Phi} )=&\;\left(\prod_j^r \pm i
\omega_j\right)\mathcal M^{0}(\phi_{1},\dots,\phi_{r},
\mathbf{\Phi} ) \notag
\\& \qquad +\mathcal{O}\left(\frac{m}{\sqrt s}\right) +\mathcal O
\left(\frac{\sqrt{s}}{\Lambda}\right)^{N+1} +\mathcal{O}\left(g,
g'\right) \quad , 
\label{eq:gbeteft} 
\end{align}
where $\mathcal{M}^{0}$ is the leading order of the matrix element in $g,g'$, and $\mathcal{O}\left(g, g'\right)$ denotes terms, which are suppressed by $g,g'$ in comparison to this leading term. 
The $\omega_j$ are specific phases that differ between initial and final states within the amplitude. 
In this framework, the matrix elements appear not only as series expansions in the gauge couplings, but also in $\sqrt s/\Lambda$, which are usually truncated after some finite order $N$. The high-energy scale $\Lambda$ of any such bottom-up EFT corresponds to a specific scale of BSM models, e.g. a reference mass of a single heavy new particle. All longitudinal gauge bosons $V_i^L$ can be replaced by the
corresponding Goldstone bosons $\phi_i$  at high energies within
the accuracy goal of the EFT. The results will match at the leading order in $g$ and $g'$. 

In the present context, we
can rewrite a modified muon Yukawa coupling as a gauge-invariant operator in
the HEFT Lagrangian, and conclude that this new interaction should cause extra
contributions to the production of multiple vector bosons in association with
the Higgs boson which rise with energy.  By construction, these contributions
exactly reproduce the effect of spoiled gauge cancellations in unitary gauge,
as computed by automated programs.

In the non-linear representation we introduce a field $U$ 
\begin{equation}
     U=e^{i\phi^a\tau_a/v} \quad \text{with} \quad \phi^a \tau_a=\sqrt{2}\begin{pmatrix}
       \frac{\phi^0}{\sqrt 2} &  \phi^+\\
       \phi^-& -\frac{\phi^0}{\sqrt 2} 
     \end{pmatrix}\quad ,
   \end{equation}
and its covariant derivative 
\begin{equation}
    D_{\mu}U=\partial_{\mu}U+igW_{\mu}U-i\frac{g'}{2}B_{\mu}U\tau_3 \quad \text{with} \quad  W_{\mu}=\frac{1}{2}\tau_a W^a_{\mu}\quad ,
\end{equation}
where $\tau_a$ denote the usual Pauli matrices and $\{\phi^+,\phi^-,\phi^0\}$ are the Goldstone bosons to the corresponding gauge bosons $\{W^+,W^-,Z\}$. The most general extension of the SM Lagrangian can be written as
\begin{align}
   \begin{split}
    \mathcal{L}_{\text{EW}}=&-\frac{1}{2} \operatorname{tr}{W_{\mu \nu}  W^{\mu
      \nu}}
      -\frac{1}{4}B_{\mu \nu}   B^{\mu \nu} + \sum_{f\in\{\ell_L,\ell_R\}} i \bar f^i
    \slashed D f^i \\
    & \qquad +\mathcal{L}_{UH}+\mathcal{L}_{\text{gauge-fix}}  \quad .
\end{split}
\end{align}
The Higgs and Goldstone sector is given by
 \begin{align}
   \begin{split}
     \mathcal L_{UH}&=\frac{v^2}{4}\operatorname{tr}[D_{\mu}U^{\dagger}D^{\mu}U] F_U(H)+\frac{1}{2}\partial_{\mu}H\partial^{\mu}H-V(H) \\
     &\qquad -\frac{v}{2\sqrt{2}}\left[\bar \ell^i_L \tilde Y_\ell^{ij}(H) U(1-\tau_3)\ell^j_R+\text{h.c.}\right] \quad,
   \end{split}
   \end{align}
where we defined the right-handed doublets as $\ell^i_R=(\nu^i_R,e^i_R)^T$, and
$i,j$ are the lepton-flavor indices.
In the SM, the functions $F_U(H), \, V(H) $ and $Y^{ij}_e(H)$ are simple polynomials in $H/v$ that can be generalized to
\begin{align}
    F_U(H)&=1+\sum_{n\geq1}f_{U,n}\left(\frac{H}{v}\right)^n ,\\  V(H)&=v^4\sum_{n\geq2}f_{V,n}\left(\frac{H}{v}\right)^n \qquad \text{and}\\
   \tilde Y_\ell^{ij}(H)&=\sum_{n\geq0}\tilde Y^{ij}_{\ell,n}\left(\frac{H}{v}\right)^n \ .
\end{align}
We do not assume CP violation in this sector,  hence the coefficient of these different series are real, 
$\tilde f_{U,n},f_{V,n},\tilde Y^{ij}_{\ell,n}\in \mathbb R$. They are general parameters
that can be obtained by a matching procedure from a possible underlying
physical model, and in principle can be measured in appropriate physical processes. 

We are primarily interested in the Higgs-lepton couplings. So we read off the mass matrix for the leptons
\begin{equation}
\tilde   M_\ell^{ij}=\frac{v}{\sqrt{2}}\tilde Y_{\ell,0}^{ij} \quad, 
\end{equation}
which is non-diagonal in general. As its eigenvalues are assumed to be positive,
we can perform the usual polar decomposition $\tilde M_\ell =U_L M_\ell
U_R^{\dagger} $ with some unitary matrices $U_{L/R}$ and compensate
this by the rotation to the physical fields $\ell_L \mapsto U_L \ell_L$ and
$\ell_R \mapsto U_R \ell_R$. 
Furthermore this defines $Y_{\ell,n} = U_L^\dagger \tilde Y_{\ell,n} U_R$, where, again, $n+1$ is the number of Higgs fields involved in the corresponding vertex. We
will focus on the physical basis from now on. 
Note, that these equations all are still matrix equations, with the (2,2)-components  $Y^{2,2}_{\ell,0}:=y_{\mu},\,Y^{2,2}_{\ell,n}:=y_{n}$ and $M^{2,2}_{\ell}:=m_{\mu}$ denoting the muon. 
Selecting the muon term and requiring the physical muon mass to equal its
observed value, we observe an effective correction of the observable Yukawa
coupling by the factor
\begin{equation}\label{eq:kmu_heft}
  \kappa_\mu = \frac{v}{\sqrt{2}m_{\mu}}y_{1}, 
\end{equation}
which, for $y_1=y_0=y_\mu$, would  correspond to the SM case $\kappa_\mu=1$.
%
A priori, the size of the coupling coefficients is unknown as it depends on the underlying dynamics. From the ``naive dimensional analysis'' \cite{Manohar:1983md,Cohen:1997rt},  one would expect the modification as $y_{n}\sim y_\mu(g^2/16\pi^2)^n$, with $g\sim 1$ for a weakly coupled theory and $g\sim {\cal{O}}(4\pi)$ a strongly coupled theory.

New operators in the series expansion in $H/v$  introduce contact terms which couple the muon to $n$ Higgs or Goldstone bosons. These contact terms are proportional to $y_m$, where $m\le n$ denotes the number of Higgs bosons and they are the leading contributions to $\mu^+\mu^-\rightarrow n \varphi$ scattering in the high energy limit.  Hence, via the GBET, a modification of $y_\mu$ is generically accompanied by new large contributions to multi-boson production in the high-energy limit.

\subsubsection{The Yukawa interaction in the SMEFT parameterization}
\label{sec:smeft}

In the SMEFT framework, the SM gauge invariance is
represented in linear form, and the Higgs boson combines with the Goldstone
bosons as a complex $SU(2)$ doublet.  The pure effect of a modified muon
Yukawa coupling can be reproduced by an infinite series of higher-dimensional
operators in the SMEFT Lagrangian~\cite{Weinberg:1979sa,Abbott:1980zj,Buchmuller:1985jz,Grzadkowski:2010es}, where all coefficients are related to the
original coupling modification.  The results will be again identical to the
unitary-gauge calculation.  

However, if we furthermore assume a
\emph{decoupling} property of the new interactions, {\it i.e.}, their parameters are not intrinsically tied to the electroweak scale, we should expect
higher-order terms in the SMEFT series to be suppressed by a new heavy physics scale  $v^2/\Lambda^2$, such that truncation
after the first term is permissible.  In that case, we have to discard the
former relation between all orders, and accept that the resulting amplitudes
will differ from the unitary-gauge results for an anomalous Yukawa
coupling.  In concrete terms, in a decoupling new-physics scenario we expect
anomalous production of multiple vector bosons to be accompanied by anomalous
production of multiple Higgs bosons.  The clean environment of a muon collider
is optimally suited to separate such final states irrespective of their decay
modes, and thus to guide model building in either direction, depending on the
pattern actually observed in data. The formalism set up here is very similar to the one used in~\cite{Falkowski:2020znk} for searching deviations in the charm and strange Yukawa couplings in multi-boson production at the LHC and FCC-hh. 

In the linear representation of the Higgs
doublet, 
\begin{equation}
  \varphi=\frac{1}{\sqrt{2}}\begin{pmatrix}
    \sqrt 2 \phi^+\\
    v+H+i \phi^0
  \end{pmatrix}\quad ,
\end{equation}
the most general bottom-up extension of the SM Lagrangian,
\begin{align}
  \begin{split}
    \mathcal{L}_{\text{EW}}=&-\frac{1}{2} \operatorname{tr}{W_{\mu \nu}  W^{\mu
      \nu}}-\frac{1}{4}B_{\mu \nu}   B^{\mu \nu} + (D_\mu
    \varphi)^{\dagger}(D^\mu \varphi)+\mu^2
    \varphi^{\dagger}\varphi-\frac{\lambda}{2}(
    \varphi^{\dagger}\varphi)^2\\ &+ \sum_{f\in\{\ell_L,e_R\}} i \bar f^i
    \slashed D f^i 
-\left(\bar \ell_L^i \tilde Y_{\ell}^{ij} \varphi e_R^j + \text{h.c.} \right)
+ \mathcal{L}_{\text{gauge-fix}} 
\end{split}
\end{align}
that leads to a modification of the Yukawa coupling, reads
\begin{equation}\label{eq:EFT}
  \mathcal L =\mathcal L_{\text{EW}}+\left [ \sum_{n=1}^N \frac{\tilde
      C^{(n)ij}_{\ell\varphi}}{\Lambda^{2n}}(\varphi^{\dagger}\varphi)^n{\bar\ell}^i_L \varphi {e^j}_R + \text{h.c.}\right ] \quad. 
\end{equation}
Operators of higher mass dimension are as usual suppressed by a large scale $\Lambda$ that can be understood as an energy cutoff for the validity of the theory, as it will lead to an expansion of the scattering matrix elements in ${\sqrt s}/{\Lambda}$. Again, we do not consider
CP violation,  hence the Wilson coefficients are real
$\tilde C^{(n)}_{\ell\varphi}\in \mathbb R$. They can be obtained by a matching procedure from an underlying
physical model, and in principle can be measured.\footnote{One rather
measures form factors, which are linear combinations of the Wilson coefficients.} For further calculations, we absorb the large scale $1/\Lambda^2$ in the Wilson coefficients. 

We can read off the (non-diagonal) mass matrix for the charged leptons
\begin{equation}
\tilde   M_\ell^{ij}=\frac{v}{\sqrt 2}\left(\tilde Y_{\ell}^{ij}-\sum_{n=1}^N
\tilde C^{(n)ij}_{\ell\varphi} \frac{v^{2n}}{2^n}\right) \quad. 
\end{equation}
In the same way as for the non-linear representation, we can diagonalize the mass matrix by redefinitions of 
the physical fields $e_L \mapsto U_L e_L$, $e_R \mapsto U_R e_R$. This defines $Y_\ell=U_L^{\dagger} \tilde Y_\ell U_R$ and $C^{(n)}_{\ell\varphi}=U_L^{\dagger} \tilde C^{(n)}_{\ell\varphi} U_R$. 

As already discussed for the non-linear case, the operator coefficients $C^{(n)}_{\ell\varphi}$ can shift the muon Yukawa coupling away from its SM value. Because of its intrinsically small value, a moderate new physics contribution could lead to a drastic effect, driving it to zero or reversing its sign. The extreme case of a vanishing muon Yukawa coupling has the significant consequence that multi-Higgs production, $\mu^+\mu^-\rightarrow H^M$ would be absent at tree level, while production of up to 
$k\in\{1,\dots,M-1\}$ Higgs bosons associated with $M-k$ vector bosons would be allowed. As a paradigm example, we show how to embed this in our SMEFT framework: we require all lepton couplings to $k$ Higgs bosons, $\Lambda_{(k)}$, $k\in\{1,\dots,M-1\}$, to vanish while the mass of the measured muon mass $m_{\mu}$ is fixed as an input. This leads to the conditions
\begin{align}
    \label{eq:SMEFT_L}
  M_\ell &=\frac{v}{\sqrt 2}\left[ Y_\ell-\sum_{n=1}^{M-1} C^{(n)}_{\ell\varphi} \frac{v^{2n}}{2^n} \right] \quad ,\\
  \Lambda_{(k)}& :=-i\frac{k!}{\sqrt 2}\left[Y_\ell\delta_{k,1}-\sum_{n=n_k}^{M-1} C^{(n)}_{\ell\varphi} \begin{pmatrix} 2n+1\\ k
  \end{pmatrix} \frac{v^{2n+1-k}}{2^n} \right] 
  = 0 
  \quad ,
\end{align}
where $n_k=\operatorname{max}(1,\lceil \frac{k-1}{2}\rceil)$. 

For the general case, we define the following modification of the SM Yukawa coupling, still matrix-valued in flavor space, as
\begin{equation}
    K_\ell =1-\frac{v}{\sqrt 2} M_\ell^{-1}\sum_{n=1}^{M-1} C^{(n)}_{\ell\varphi} \frac{n v^{2n}}{2^{n-1}} \quad .
\end{equation}
Again, we can project to the muon via $Y^{2,2}_{\ell}:=y_{\mu},\,C^{(n)2,2}_{\ell\varphi}:=c^{(n)}_{\ell\varphi},   M^{2,2}_{\ell}:=m_{\mu}$, as well as 
$K^{2,2}_{\ell}:=\kappa_{\mu}$.

As usual, we will consider the linear SMEFT expansion up to the first non-trivial order, which adds to the dimension-4
SM Yukawa coupling operator, $\mathcal{L}_{\text{Yuk.}} \,  = \; -(\bar\ell_L Y_\ell e_R)\varphi$ at dimension-6 
a single operator that modifies the static Higgs coupling  to leptons:
\begin{equation}
  \label{eq:O_ephi-6}
  \mathcal{O}_{\ell\varphi} =
  C_{\ell\varphi}(\varphi^\dagger\varphi)(\bar\ell_L
  e_R)\varphi \ .
\end{equation}
Here, both $\Gamma_\ell$ as well as $C_{\ell\varphi}$
are matrices in lepton-flavor space. On dimensional grounds, $C_{\ell\varphi}\sim 1/\Lambda^2$, where $\Lambda$ is the scale at which new physics sets in.
Inserting the Higgs vev, we obtain at dimension-4 the SM value of the lepton mass matrix, $M_\ell^{(4)} = \frac{v}{\sqrt2}Y_\ell$, while at dimension-6 we get 
a modified mass matrix
\begin{equation}
  M_\ell^{(6)} = \frac{v}{\sqrt2}\left(Y_\ell - \frac{v^2}{2}C_{\ell\varphi}\right) .
\end{equation}
Specializing to the muon term and requiring the physical muon mass to equal its measured value, we observe an effective modification of the observable Yukawa
coupling by the factor
\begin{equation}
\label{eq:kmu_smeft}
  \kappa_\mu^{(6)} = 1 - \frac{v^3}{\sqrt2\,m_\mu}c_{\ell\varphi}^{(1)}.
\end{equation}
Expanding the Higgs field, the new operator induces contact terms which couple the muon to $n=1, 2$, or 3 Higgs or Goldstone bosons.  The contact terms are
all proportional to the operator coefficient  $c_{\ell\varphi}^{(1)}$, either scalar or
pseudoscalar.  Squaring this interaction, we obtain local contributions to $\mu^+\mu^-\to n\varphi$ scattering, in analogy with the HEFT description.
The physical final states are Higgs or longitudinal $W,Z$ gauge bosons.  As we will discuss in more detail in Sec.~\ref{sec:ratios}, the $d=6$ contributions to their production cross sections with multiplicity $n=3$
rise with energy, $\sigma \propto s$, 
while the SM contribution falls off like $1/s$. There is no interference, since -- for these final states -- the SM requires a vector exchange while the new contact
term is scalar.  We obtain a deviation from the SM prediction which is
determined by the EFT contribution alone, which becomes leading above some
threshold which depends on $\kappa^{(6)}_\mu-1$.  The decomposition of the anomalous
contribution into particle types ($WWZ$, $WWh$, etc.) is fixed by electroweak
symmetry and the particular SMEFT operator content, such that the exclusive
channels are related by simple rational factors beyond the threshold where the
new-physics part starts to dominate the production rates. This will be elaborated in Sec.~\ref{sec:ratios}.

If the correction was large enough to render $\kappa_\mu=0$, we would obtain the unitarity bound for $d=6$, {\it i.e.}~three-boson emission, as discussed in the next subsection.
Generally speaking, the modification from the SM Yukawa coupling could reach an order of $100\%$ if  $c_{\ell\varphi}^{(1)} \sim 0.1/(10v)^2$.
We emphasize that these two sample scenarios -- a pure modified Yukawa coupling, and a modified Yukawa coupling combined with truncation of the SMEFT series -- are to be understood as mere representatives of a potential new class of SM modifications that are difficult to observe at lower energy.  As
our results indicate, there is a great redundancy in the analysis of exclusive multi-boson final states, which should translate into significant discrimination power regarding more detailed models of the Higgs-Yukawa sector beyond the SM.
If we translate an experimental bound on $\Delta\kappa_\mu$ to the SMEFT coefficient $c^{(1)}\sim g/\Lambda^2,$ we obtain a bound on the scale of new physics as 
\begin{equation}
\Lambda >10\ {\rm TeV}\sqrt{\frac{g}{\Delta\kappa_\mu}}\quad.
\label{eq:bound} 
\end{equation}

\subsubsection{Unitarity bounds on a nonstandard Yukawa sector}
\label{sec:unitarity}

In the SM, the high-energy asymptotics of the multi-boson production cross sections  universally fall off with rising energy, manifesting themselves in delicate gauge
cancellations which become huge at high energies. A modification of the
muon Yukawa coupling from the SM prediction would show up as spoiling such cancellations, and thus 
eventually causes specific scattering amplitudes to rise again, without
limits.  While in theory, such a unitary-gauge framework does not do justice to
the built-in symmetries of the SM, it is nevertheless the baseline framework for any tree-level evaluations
such as the ones that we use in this work.

In Ref.~\cite{Maltoni:2001dc}, generic models have been 
investigated where the leading contribution to a fermion mass
originates from a dimension-$d$ EFT operator that couples the fermion to the
SM Higgs field.  Using the GBET, they computed the
energy scale $\Lambda_d$ where unitarity is violated by multiple emission of
Goldstone bosons, representing longitudinally polarized weak vector bosons,
and Higgses.
\begin{equation}
  \Lambda_d = 4\pi\kappa_d\left(\frac{v^{d-3}}{m_f}\right)^{1/(d-4)},
  \quad\text{where}\quad
  \kappa_d = \left(\frac{(d-5)!}{2^{d-5}(d-3)}\right)^{1/(2(d-4))}.
\end{equation}
For any given $d>4$, the most relevant bound corresponds to a final state that
consists of $n=d-3$ Goldstone or Higgs bosons in total.  For $m_f=m_\mu$ and
$d=6,8,10$, the numeric values of the unitarity bound are $95\,\TeV$,
$17\,\TeV$, and $11\,\TeV$, respectively.  For $d\geq 8$, the values of these
bounds lie within the energy range that is accessible at a future muon
collider.  They imply large amounts of observable multi-boson production.  The
strong suppression of the corresponding SM processes enables a study already
significantly below those upper bounds.  Furthermore, we expect observable
effects even if only a fraction of the muon mass is due to the new-physics
contributions that are parameterized by those operators.

In the previous subsection, we have discussed an analogous sequence of phenomenological scenarios within the SMEFT framework, where we require that local Higgs-fermion couplings are absent up to a given Higgs multiplicity $n$.  This requirement enforces a specific choice of the SMEFT operator coefficients $C_{\ell\varphi}^{(n)}$ up to dimension $d=2n+4$, as defined by~\eqref{eq:SMEFT_L}. The limit $d\to\infty$ corresponds to the case of no local Higgs-fermion couplings of any multiplicity.  We emphasize that this peculiar choice is merely an extreme case of a generic anomalous muon Yukawa sector.  The generic case is parameterized within the SMEFT or HEFT formalisms, allowing the coefficients of the higher-dimensional couplings to vary freely within the constraints imposed by unitarity.

\begin{figure}
\centering
\includegraphics[width=\textwidth]{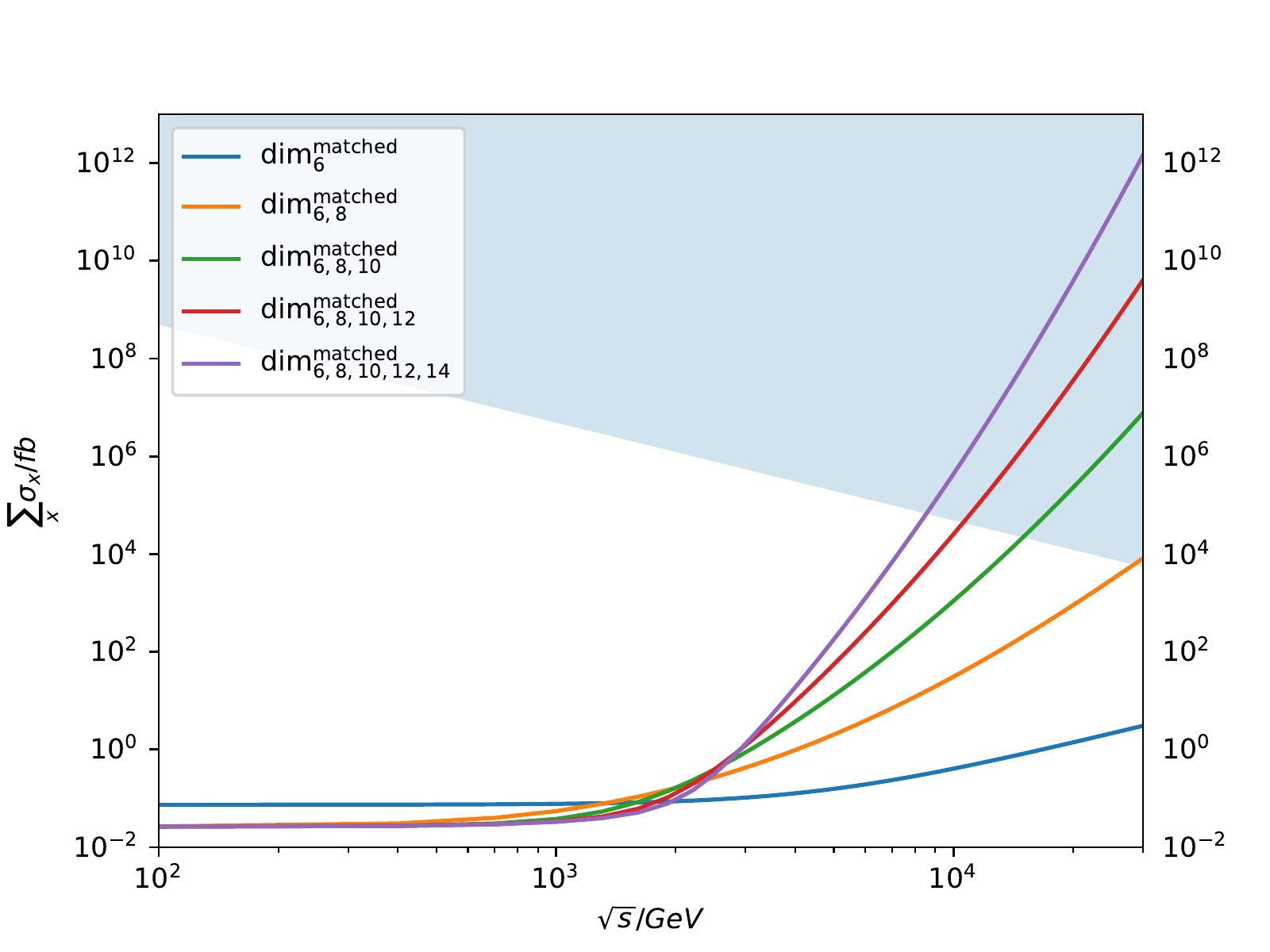}
    \hspace*{4cm}
    \caption{Inclusive inelastic cross section $\mu^+\mu^-\to X$ for multiple Goldstone and Higgs-boson production in the GBET approximation.  We show the result for the sequence of SMEFT scenarios defined by the conditions~\eqref{eq:SMEFT_L}, truncated at dimension $d=6,8,10,12,14$, respectively.  The maximal multiplicity of the final state is $n=3,5,7,9,11$, respectively.  The shaded area indicates the region that is excluded by the universal unitarity bound for the inclusive cross section~\eqref{eq:ubound}. 
    } 
\label{fig:unitarity}
\end{figure}

In quantitative terms, the unitarity constraint for the total inelastic cross section $\sigma_{\mu^+\mu^-\to X}(s)$, where $X\neq\mu^+\mu^-$, is given by the inequality
\begin{equation}
    \label{eq:ubound}
    \sum_X \sigma_{\mu^+\mu^-\to X}(s) \leq \frac{4\pi}{s}.
\end{equation}
In Fig.~\ref{fig:unitarity} we display the total cross section for this sequence of scenarios, including operators up to dimension $d=6,8,10,\dots$ and compare it with the upper bound~\eqref{eq:ubound}.  The cross section has been evaluated using the GBET, summing over all final states.  The SM contribution ($d=4$) can be neglected for this purpose, and the boson masses are set to zero.  The multiplicity of the Higgs and Goldstone bosons extends up to $n=d-3$, which evaluates to $n=3,5,7,\dots$, respectively.

We observe that for $d\leq 10$ (i.e., $n\leq 7$), the sum over cross sections does not touch the unitarity bound before $15\,\TeV$, while for higher dimension and multiplicity, the curves cross already at collider energies within the range considered for a muon collider.
In the $d\to\infty$ case, the multiplicity of extra Goldstone-boson production
becomes unbounded, and the unitarity limit for the sequence of scenarios~\eqref{eq:SMEFT_L} formally drops towards the original
electroweak scale~\cite{Maltoni:2001dc}.  Even if we account for finite
vector-boson masses, such a scenario should be qualified as strongly
interacting, and finite-order predictions in the multi-TeV range become
invalid. Of course, we do not expect the actual operator coefficients to strictly follow such a pattern, so the argument should rather be understood as a guideline regarding the inherent limitations of the EFT in the current context.

For this reason, we consider lower-dimensional operators in the
SMEFT or HEFT expansions individually.  The presence of extra Higgs bosons in the
gauge-invariant SMEFT operators of fixed dimension delays the potential onset
of new (strong) interactions to higher energy. While in the tables and plots of the subsequent sections we will frequently refer to the $d=\infty$ limit for illustration, in our phenomenological study we work with Higgs--Goldstone multiplicities $n\leq 4$ and limit the dimensions of the included SMEFT operators to $d=6,8,10$.  For those final states, Fig.~\ref{fig:unitarity} indicates that unitarity is not yet relevant at a muon collider as proposed, even if we adopt one of the extreme scenarios described above.  Clearly, higher multiplicities may yield even stronger effects, but their contributions depend on further coefficients in the EFT expansion and should therefore be regarded as model-dependent.  In fact, if in~\eqref{eq:ubound} we restrict the sum over final states to $n\leq 4$, there is no problem with unitarity for any of the parameter sets shown in Fig.~\ref{fig:unitarity}.  The numerical results of our study below will rely on the lowest multiplicities and analyze small deviations from the SM where the actual effect is at the limit of the collider sensitivity, orders of magnitude below the unitarity bound.

\subsubsection{Multi-boson production and cross section ratios}
\label{sec:ratios}

Obviously, the most direct and model-independent probe to the muon-Higgs coupling would be the $s$-channel resonant production 
$$\mu^+\mu^-\to H.$$
This was the motivation for a muon-collider Higgs factory~\cite{Barger:1995hr,Barger:1996jm}. 
This process would put an extremely high demand on the collider beam quality to resolve the narrow width of the Higgs boson, and on the integrated luminosity. Off the resonance at higher energies, one could consider to study this coupling by utilizing the process of radiative return~\cite{Chakrabarty:2014pja}. 
Although the expected cross sections for multiple Higgs production $\mu^+\mu^- \to HH$ and $HHH$ 
are quite small as shown later, they receive 
a power enhancement $E/\Lambda$ of the effective coupling of $\kappa_\mu$, 
if a new interaction like the dimension-6 operator, Eq.~\eqref{eq:O_ephi-6}, 
is present.
If an analogous dimension-8 operator is present with a Wilson coefficient $c_{\ell\varphi}^{(2)}\sim 1/\Lambda^4$, the physical muon mass and the Yukawa couplings are given by

\begin{align}
  \label{eq:d=6+8-m}
  m_\mu^{(8)} &= \frac{v}{\sqrt2}\left(y_{\mu} -
  \frac{v^2}{2}c^{(1)}_{\ell\varphi} - \frac{v^4}{4}c^{(2)}_{\ell\varphi}\right),
  \\
  \label{eq:d=6+8-l}
  \lambda_\mu^{(8)} &= \phantom{\frac{v}{\sqrt2}}
  \left(y_{\mu} -
  \frac{3v^2}{2}c^{(1)}_{\ell\varphi} - 
  \frac{5v^4}{4}c^{(2)}_{\ell\varphi}\right),
\end{align}
The dimension-8 operator causes a rise of $n$-boson production cross sections, and ultimately a saturation of tree-level unitarity, for up to $n=5$  as discussed in the previous section. Depending on the relative size of the individual contributions at a 
given energy, the ratios of individual multi-boson channels are determined by
either $Y_e$, $C^{(1)}_{\ell\varphi}$ or $C^{(2)}_{\ell\varphi}$ . 
Final states with more Higgs bosons receive direct contributions which rapidly rise with energy $(E/\Lambda)^n$.

The operators introduced in Eqs.~\eqref{eq:EFT} and \eqref{eq:d=6+8-m}$-$\eqref{eq:d=6+8-l} induce contact terms, schematically written as,
\begin{center}
\raisebox{-.45\height}{
\includegraphics[width=.25\textwidth]{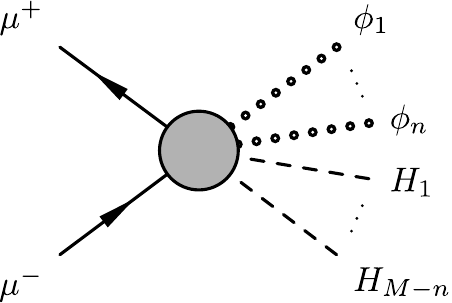}
}
$\approx$
\raisebox{-.45\height}{
\includegraphics[width=.25\textwidth]{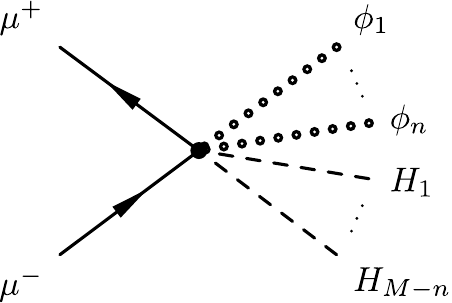}
}
\end{center}
which are dominant in the high-energy limit as there is no suppression in $\sqrt{s}$ from propagator denominators. Let us denote the Feynman rules for a multi-boson final state $X$ as
\begin{center}
$\left. 
\raisebox{-.45\height}{
\includegraphics[width=.25\textwidth]{figs/muon_multiboson_contact}
} \qquad
\right\}  \quad 
X_i : \qquad i \;C_{X_i} (P_L \pm P_R) \quad ,$
\end{center}
where $C_{X_i}$ is a linear combination of Wilson coefficients, and $i$ labels all possible final states for a given multiplicity. The sign in $(P_L \pm P_R)$ depends on the number of Goldstone bosons $\phi^0$ in the final state and does not play any role for the following argument.  The spin-averaged matrix element reads
($k_i, i=1,2$ are the two muon momenta, $s=2 k_1\cdot k_2$, where we ignored the muon mass in the kinematics of the matrix element)
\begin{align*}
  \overline {|\mathcal A_{X_i}|^2}
  &=\frac{1}{4} |C_{X_i}|^2\sum_{s_1,s_2} \bar v_{s_1}(k_1)(P_L\pm
  P_R)u_{s_2}(k_2)\bar u_{s_2}(k_2)(P_R\pm P_L)v_{s_1}(k_1)
  \\ 
  & 
  =|C_{X_i}|^2 \times ( k_1 \cdot k_2 \mp m_{\mu}^2)
   \approx \ \frac{|C_{X_i}|^2 s}{2} \quad .
\end{align*}
As the spin-averaged matrix element in that approximation is constant, the integration over 
the phase space is trivial and yields a cross section
\begin{equation}
  \sigma^{X_i}=\frac{(2\pi)^4}
  {2s} \; 
  |\mathcal A_{X_i}|^2 \; \left(\prod_{j\in
    J_{X_i}}\frac{1}{n_j !}\right)\,\Phi_{M}(k_1+k_2;p_1,\dots ,p_{M})
  \quad , 
\end{equation}
where $\Phi^{X_i}_{M}(k_1+k_2;p_1,\dots ,p_{M})$ is the
$M$-particle phase-space volume and $J_{X_i}$ is the set of
indistinguishable particles $X_i$ in the final state with numbers $n_j$ for particle $j\in J_{X^i}$. 
As we study the limit of very high energies, we 
neglect all particle masses, and the phase-space volume will be the same for all final states $X_i$. In the center-of-mass (CMS) system (cf.~\cite{Kleiss:1985gy}), the $M$-particle phase space is given by ($\Gamma$ is the Euler gamma function)
\begin{align}
  \Phi^{X_i}_{M}(k_1+k_2;p_1,\dots
  ,p_{M})=\frac{1}{(2\pi)^{3M}}\left(\frac{\pi}{2}\right)^{M-1}\frac{s^{M-2}}{\Gamma(M)
    \Gamma(M-1)} \quad .
\end{align}
In order to study the effects from specific operator coefficients, it is beneficial to look into ratios of cross sections with respect to a certain reference cross section for a specific exclusive final state of the same multiplicity. For such cross-section ratios we find
\begin{equation}
  R^{X_i}:=\frac{\sigma^{X_i}}{\sigma^{X_{\text{ref}}}}=\frac{|C_{X_i}|^2\left(\prod_{j\in
      J_{X_i}}\frac{1}{n_j
      !}\right)}{|C_{X_{\text{ref}}}|^2\left(\prod_{j\in
      J_{X_{\text{ref}}}}\frac{1}{n_j !}\right)} \quad. 
\end{equation}

  \begin{table}
   \begin{center}
   \begin{tabular}{ c||c|c|c|c||c|c}
     \hline
   & \multicolumn{6}{|c}{$\Delta\sigma^{X}/\Delta\sigma^{W^+W^-}$}\\
   \hline       
    &   
       \multicolumn{4}{|c||}{SMEFT} &
       \multicolumn{2}{c}{HEFT}
       \\
     \hline
   $X$ &dim$_6$ & dim$_8$ & dim$_{6,8}$ & dim$_{6,8}^{\text{matched}}$ & dim$_\infty$ & dim$_\infty^{\text{matched}}$\\
   \hline
   $W^+W^-$ & $1$ & $1$ & $1$ & $1$ & $1$ & $1 $\\
   $ZZ$ & $1/2$ & $1/2$ & $1/2$ & $1/2$ & $1/2$ & $1/2$\\
    \hline
   $ZH$ & $1$ & $1/2$ & $1$ & $1$ & $R^{\text{HEFT}}_{(2),1}$ & $1$\\
   $HH$ & $9/2$ & $25/2$ & $ R^{\text{SMEFT}}_{(2),1}/2$ & $0$ & $2\, R^{\text{HEFT}}_{(2),2}$ & $0$\\
   \hline
   \end{tabular}
   \end{center}
    \caption{Ratios of final-state cross-section deviations in diboson production, assuming that the leading muon-Yukawa   contribution originates from various combinations of $d=6$   and $d=8$ operators in SMEFT, or from a direct contribution in the HEFT, respectively. The
    term ``matched"  indicates the matching to a model with a vanishing muon-Yukawa coupling. See the text for details. The coefficients $R_{(2),i}$ are defined in~\eqref{eq:Rin2}.}
    \label{tab:ratios-2}
  \end{table}

In the following, we discuss ratios of deviations of production cross sections from their SM values for final-state multiplicities $n=2,3,4$. For each multiplicity, the cross-section deviations $\Delta\sigma^X$ for different final states $X$ will be normalized with respect to a particular exclusive reference final state, which is $W^+W^-$ for dibosons, $W^+W^-H$ for tribosons, and $W^+W^-HH$ for four bosons, respectively. The cross sections are calculated in the GBET approximation for massless Goldstone bosons; for longitudinal $W^\pm$ and $Z$ boson final states they become exact in the limit that both their masses as well as the SM contributions to these cross sections can be neglected. We are considering these ratios for different EFT scenarios, namely for truncating the SMEFT series of higher-dimensional operators at dimension $d=6,8,10$, respectively, as well as for the non-linear HEFT case. 

In detail, in Table~\ref{tab:ratios-2} we consider the diboson final states for the cases of a pure $d=6$ contribution (dim$_6$), a pure $=8$ contribution (dim$_8$), a mixed contribution (dim$_{6,8}$), and for the case where the $d=6$ and $d=8$ operators are tuned to cancel the leading-order Yukawa coupling according to~\eqref{eq:d=6+8-m}, \eqref{eq:d=6+8-l}, denoted dim$_{6,8}^\text{matched}$. For the non-linear HEFT setup, the first column (dim$_\infty$) takes into account the full tower, in principle, though only the lowest dimension contributes at tree level due to the $n$-arity of the vertex. The last column (dim$_\infty^\text{matched}$) is the matched case again with a vanishing Yukawa coupling, calculated by taking into account a sufficiently large number of terms corresponding to the linear setup. The list of processes includes direct production of up to two Higgs bosons.
The non-rational coefficients in this and the following tables are expressed in terms of ratio coefficients, $R^{\text{HEFT/SMEFT}}_{(N),i}$, where $N$ is the multiplicity of the boson final state, and $i$ labels the contribution from higher-dimensional operators to the given multiplicity with increasing operator order,
\begin{align}
  \label{eq:Rin2}
  R^{\text{SMEFT}}_{(2),1}&=\left(\frac{5v^2 c^{(2)}_{\ell\varphi}+c^{(1)}_{\ell\varphi}}{v^2 c^{(2)}_{\ell\varphi}+c^{(1)}_{\ell\varphi}}\right)^2,
  &
  R^{\text{HEFT}}_{(2),1}&=\left(\frac{y_{1}}{y_{\mu}}\right)^2,
  &
  R^{\text{HEFT}}_{(2),2}&=\left(\frac{y_2}{y_{\mu}}\right)^2 \ .
\end{align}
Here, the $c^{(i)}_{\ell\varphi}$ operator coefficients of SMEFT have been introduced above in~\eqref{eq:d=6+8-m}, \eqref{eq:d=6+8-l}, while by $y_i$ we have denoted the Yukawa
couplings of the muon to $i+1$ Higgs bosons in the HEFT parameterization. In SMEFT, if the dim$_6$ contributions dominate, then $R^{\rm SMEFT}\sim 1$. On the other hand, the dim$_8$ contributions can modify this behavior. In HEFT, $R^{\rm HEFT}$ could be larger than 1 in a strongly coupled theory.  In addition, those anomalous contributions will lead to enhancements at high energies.

\begin{table}
 \begin{center}
 \begin{tabular}{ c||c|c|c|c||c|c}
   \hline
     & \multicolumn{6}{|c}{$\Delta\sigma^{X}/\Delta\sigma^{W^+W^-H}$}\\
   \hline       
    &   
       \multicolumn{4}{|c||}{SMEFT} &
       \multicolumn{2}{c}{HEFT}
       \\
   \hline
 $\mm\to X$ & dim$_6$ & dim$_8$ & dim$_{6,8}$ & dim$^{\text{matched}}_{6,8}$ & dim$_\infty$ & 
 dim$^{\text{matched}}_\infty$ \\
 \hline
 $WWZ$ & $1$ & $1/9$ & $R^{\text{SMEFT}}_{(3),1}$ & $1/4$ & $ R^{\text{HEFT}}_{(3),1}$/9 & $1/4$\\
 $ZZZ$ & $3/2$ & $1/6$ & $3 \, R^{\text{SMEFT}}_{(3),1}/2$ & $3/8$ & $R^{\text{HEFT}}_{(3),1}/6 $& $3/8$ \\
  \hline
 $WWH$ & $1$ & $1$ & $1$ & $1$ & $1$ & $1$\\
 $ZZH$ & $1/2$ & $1/2$ & $1/2$ & $1/2$ & $1/2$& $1/2$\\
 $ZHH$ & $1/2$ & $1/2$ & $1/2$ & $1/2$ & $2\, R^{\text{HEFT}}_{(3),2}$ & $1/2$\\
 $HHH$ & $3/2$ & $25/6$ & $3\, R^{\text{SMEFT}}_{(3),2}/2$ & $75/8$ & $6\, R^{\text{HEFT}}_{(3),3}$ & $0$\\
 \hline
 \end{tabular}
 \end{center}
  \caption{Same as Tab.~\ref{tab:ratios-2} but for triboson production. 
  The coefficients $R_{(3),i}$ are listed in~\eqref{eq:Rin3i}-\eqref{eq:Rin3f}.
  }
  \label{tab:ratios-3}
\end{table}

\begin{table}
   \begin{center}
     \begin{tabular}{ c||c|c|c|c||c|c }
       \hline
       & \multicolumn{6}{|c}{$\Delta\sigma^{X}/\Delta\sigma^{WWHH}$}
       \\
       \hline
       & 
       \multicolumn{4}{|c||}{SMEFT} &
       \multicolumn{2}{c}{HEFT}
       \\
       \hline
       $\mm\to X$ &dim$_{6,8}$ & dim$_{10}$ & dim$_{6,8,10}$ & dim$^\text{matched}_{6,8,10}$ & dim$_\infty$ & dim$^\text{matched}_\infty$ \\
       \hline
       $WWWW$ & $2/9$ & $2/25$ & $2 \, R^{\text{SMEFT}}_{(4),1}/9$ & $1/2$ & $R^{\text{HEFT}}_{(4),1}/18$ & $1/2$\\
       $WWZZ$ & $1/9$ & $1/25$ & $ R^{\text{SMEFT}}_{(4),1}/9$ & $1/4$ & $  R^{\text{HEFT}}_{(4),1}/36$ & $1/4$\\
       $ZZZZ$ & $1/12$ & $3/100$ & $ R^{\text{SMEFT}}_{(4),1}/12$ & $3/16$ & $  R^{\text{HEFT}}_{(4),1}/48$ & $3/16$\\
       \hline
       $WWZH$ & $2/9$ & $2/25$ & $2 \, R^{\text{SMEFT}}_{(4),1} /9$ & $1/2$ & $R^{\text{HEFT}}_{(4),2}/8$ & $1/2$\\
       $WWHH$ & $1$ & $1$ & $1$ & $1$ & $1$ & $1$\\
       $ZZZH$ & $1/3$ & $3/25$ & $R^{\text{SMEFT}}_{(4),1}/3$ & $3/4$ & $R^{\text{HEFT}}_{(4),2}/12$ & $3/4$\\
       $ZZHH$ & $1/2$ & $1/2$ & $1/2$ & $1/2$ & $1/2$ & $1/2$\\
       $ZHHH$ & $1/3$ & $1/3$ & $1/3$ & $1/3$ & $3\, R^{\text{HEFT}}_{(4),3}$ & $1/3$\\
       $HHHH$ & $25/12$ & $49/12$ & $25\, R^{\text{SMEFT}}_{(4),2}/12$ & $1225/48$ & $12\, R^{\text{HEFT}}_{(4),4}$  & $0$\\
       \hline
 \end{tabular}
 \end{center}
  \caption{Same as Tabs.~\ref{tab:ratios-2} and \ref{tab:ratios-3} but for four-boson production.
    The coefficients $R_{(4),i}$ are listed in~\eqref{eq:Rin4i}-\eqref{eq:Rin4f}.}
   \label{tab:ratios-4}
\end{table}

The cross-section ratios in the case of triboson production are summarized in Table~\ref{tab:ratios-3}. Here, all exclusive final-state production cross sections are normalized to the $W^+W^-H$ final state, which is the one whose phenomenology we will study in detail in Sec.~\ref{sec:Pheno}. As for the case of diboson production, we consider scenarios with a pure $d=6$ contribution (dim$_6$), a pure $d=8$ contribution (dim$_8$), a mixed contribution (dim$_{6,8}$), and for the case where the $d=6$ and $d=8$ operators are tuned to cancel the leading-order Yukawa coupling according to~\eqref{eq:d=6+8-m}, \eqref{eq:d=6+8-l} (dim$_{6,8}^\text{matched}$), respectively. Exclusive final states contain up to three physical Higgs bosons. For the triboson case, we define the following ratio coefficients for the SMEFT and HEFT case, respectively, as 
\begin{align}
  \label{eq:Rin3i}
  R^{\text{SMEFT}}_{(3),1}&=\left(\frac{v^2 c^{(2)}_{\ell\varphi}+c^{(1)}_{\ell\varphi}}{3v^2 c^{(2)}_{\ell\varphi}+c^{(1)}_{\ell\varphi}}\right)^2,
  &
  R^{\text{SMEFT}}_{(3),2}&=\left(\frac{5v^2 c^{(2)}_{\ell\varphi}+c^{(1)}_{\ell\varphi}}{3v^2 c^{(2)}_{\ell\varphi}+c^{(1)}_{\ell\varphi}}\right)^2
\end{align}
and 
\begin{align}
  \label{eq:Rin3f}
  R^{\text{HEFT}}_{(3),1}&=\left(\frac{y_{\mu}}{y_1}\right)^2,
  &
  R^{\text{HEFT}}_{(3),2}&=\left(\frac{y_2}{y_1}\right)^2,
  &
  R^{\text{HEFT}}_{(3),3}&=\left(\frac{y_{3}}{y_1}\right)^2 \qquad .
\end{align}

We recall that at multiplicity $n=4$ and beyond,
the dimension-6 SMEFT operator does not directly contribute in the GBET approximation, so we choose to include the effects of the analogous dimension-8 and dimension-10 operators in the table for the production of quartic final states. 
In Table~\ref{tab:ratios-4}, we display the ratios of four-particle final state cross sections; definitions and conventions are analogous to those in Table~\ref{tab:ratios-3}. The ratio coefficients for the four-boson final states are given by
\begin{align}
  \label{eq:Rin4i}
  R^{\text{SMEFT}}_{(4),1}&=\left(\frac{3v^2 c^{(3)}_{\ell\varphi}+2c^{(2)}_{\ell\varphi}}{5v^2
    c^{(3)}_{\ell\varphi}+2c^{(2)}_{\ell\varphi}}\right)^2,
  &  
  R^{\text{SMEFT}}_{(4),2}&=\left(\frac{7v^2 c^{(3)}_{\ell\varphi}+2c^{(2)}_{\ell\varphi}}{5v^2
    c^{(3)}_{\ell\varphi}+2c^{(2)}_{\ell\varphi}}\right)^2
\end{align}
and 
\begin{align}
  \label{eq:Rin4f}
  R^{\text{HEFT}}_{(4),1}&=\left(\frac{y_{\mu}}{y_2}\right)^2,
  &
  R^{\text{HEFT}}_{(4),2}&=\left(\frac{y_1}{y_2}\right)^2,
  &
  R^{\text{HEFT}}_{(4),3}&=\left(\frac{y_3}{y_2}\right)^2,
  &
  R^{\text{HEFT}}_{(4),4}&=\left(\frac{y_4}{y_2}\right)^2.
\end{align}

\begin{figure}
\centering
\includegraphics[width=0.8\textwidth]{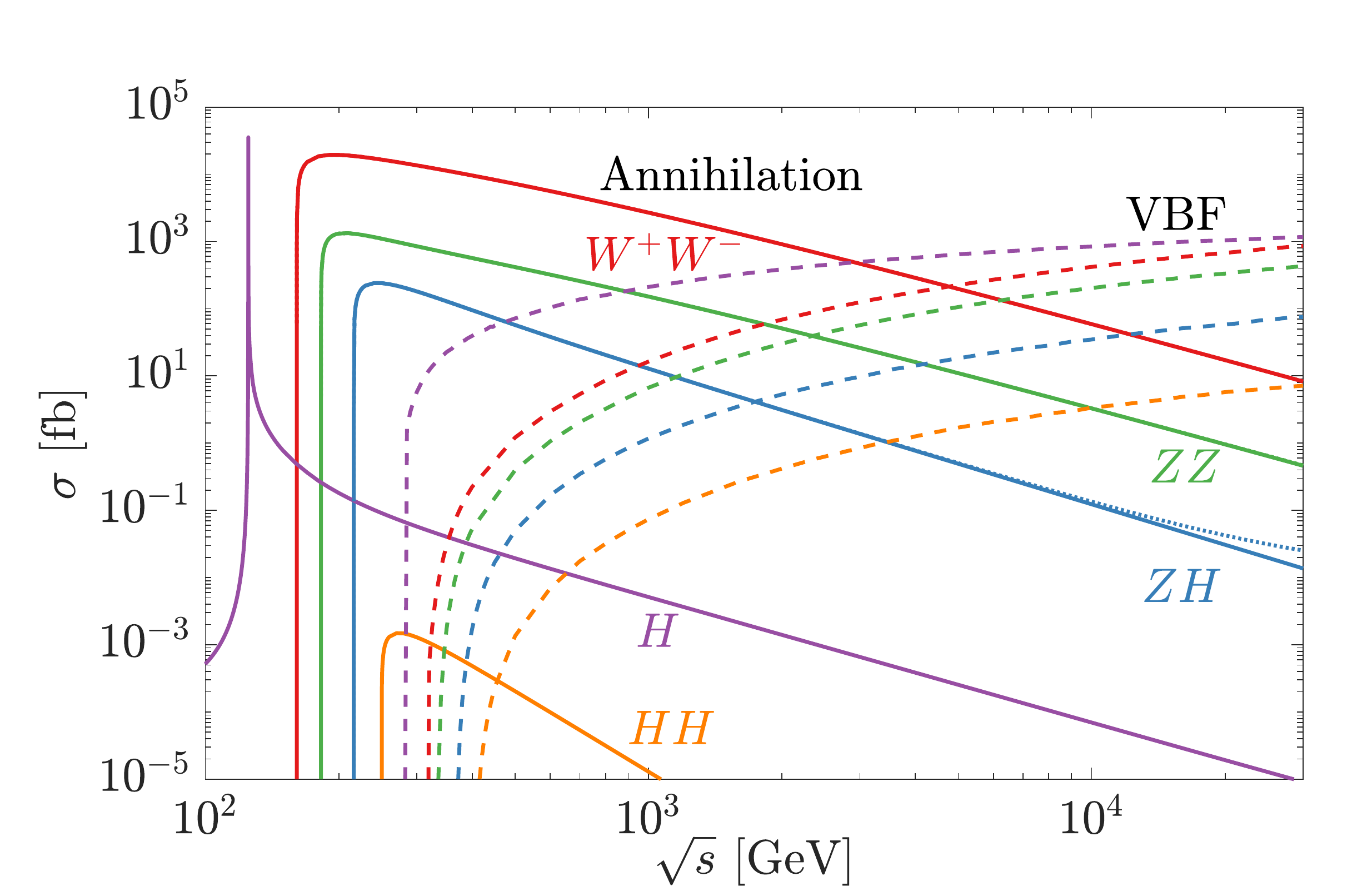}
\caption{The cross sections of diboson production at a $\mm$ collider as a function of the c.m. energy $\sqrt{s}$. The solid and dotted lines are for the direct annihilation with muon Yukawa coupling as $\kappa_\mu=1$ and $\kappa_\mu=0~(2)$ (hardly visible), respectively. The dashed rising curves are the (charged) vector boson fusions (VBF), $\mm\to\nn X$, calculated using the fixed-order (FO) approach with a cut on the invariant mass of $\nn $ pair $M_{\nn} > 150 \,{\rm GeV}$. All calculations are carried out  with {\sc Whizard~2.8.5}.} 
\label{fig:2B}
\end{figure}

To numerically cross check the analytical results for the cross-section ratios, we implemented the extreme case of the SM with a vanishing as well as with a $\kappa$-rescaled muon Yukawa coupling, respectively, within the same Monte Carlo (MC) framework that we used for our phenomenological study in Sec.~\ref{sec:Pheno} for multi-boson final states $X_i$ for the class of processes $\mu^+\mu^-\rightarrow W^+W^-H^{M-2}$. Our numerical MC results agree perfectly with the ratios given in Tables \ref{tab:ratios-2}, \ref{tab:ratios-3}, and~\ref{tab:ratios-4}, thereby validating our SMEFT implementation. 

In summary, the common feature of all versions of the modified Yukawa sector
is a proliferation of multi-boson production at high energy.  The anomalous
contributions do not interfere with SM production due to the mismatch in
helicity.  The dimensionality of the anomalous interactions determines the
particle multiplicity in the energy range where the new interactions start to
dominate over SM particle production.  The breakdown into distinct final
states allows for drawing more detailed conclusions on the operator content
and thus the underlying mechanism.

In the next section, we are studying the phenomenology of such a SMEFT setup featuring a modified muon Yukawa coupling and assess our sensitivity to it at a high-energy $\mm$ collider, using the paradigm process $\mm \to W^+W^-H$. Processes with multiple Higgs bosons only in the final state are also very interesting and may yield further strong signals, as can be read off from the tables above.  The SM rates for those final states are tiny, so any signal is a clear indication for new physics in this sector.  However, the cross sections of pure multi-Higgs final states such as $HHH$ are also more model-dependent.  By adjusting the higher-order coefficients in the SMEFT expansion, those cross sections can be varied at will without altering the ordinary muon Yukawa coupling.  This is evident since in the alternative HEFT formalism where the Higgs is a singlet, the local couplings to different numbers of Higgs bosons are not related at all, cf.\ Sec.~\ref{sec:heft}.  Turning the argument around, if an anomalous Goldstone-boson signal is found as we study below, analyzing the relative magnitude of pure-Higgs final states will reveal details about the underlying Higgs-sector dynamics.  We defer this to a separate phenomenological study.

\section{Phenomenology of Muon-Higgs  Coupling at a high-energy Muon Collider}
\label{sec:Pheno}

In this section, we explore the phenomenology of multi-boson production for the sensitivity to the muon Yukawa coupling at a muon collider with collision energy in the range $1<\sqrt{s}<30$ TeV, with an integrated luminosity, which scales with energy quadratically as
\cite{Delahaye:2019omf,Bartosik:2020xwr},
\begin{equation}\label{eq:lumi}
  \mathcal{L}=\left(\frac{\sqrt{s}}{10\TeV}\right)^2 10~\textrm{ab}^{-1}.
\end{equation}

\begin{figure}
\centering
\includegraphics[width=0.8\textwidth]{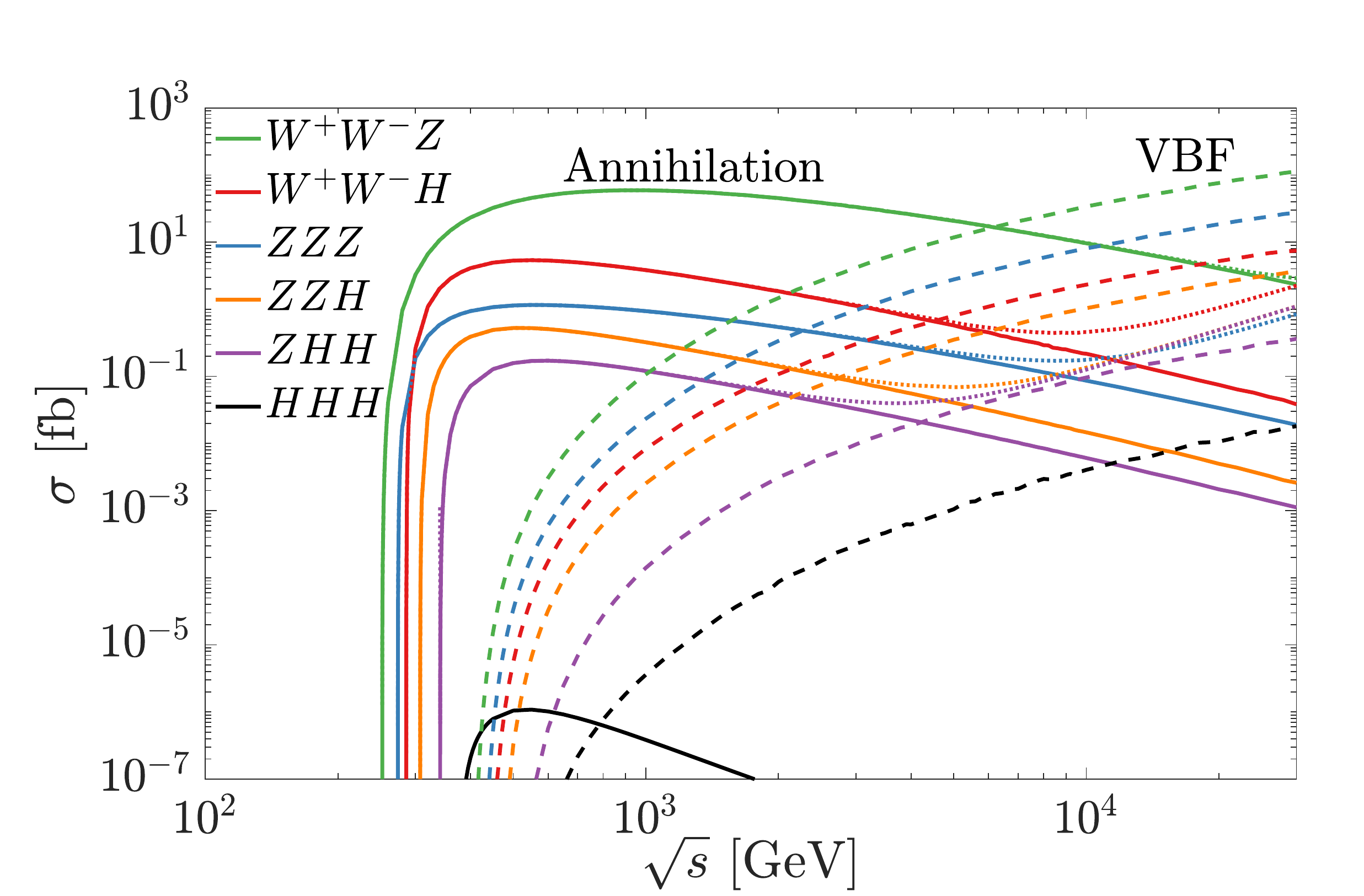}
\caption{Similar to Fig.~\ref{fig:2B}, the cross sections of three-boson production at a $\mm$ collider as a function of the c.m. energy $\sqrt{s}$.} 
\label{fig:3B}
\end{figure}

\begin{figure}
\centering
\includegraphics[width=0.8\textwidth]{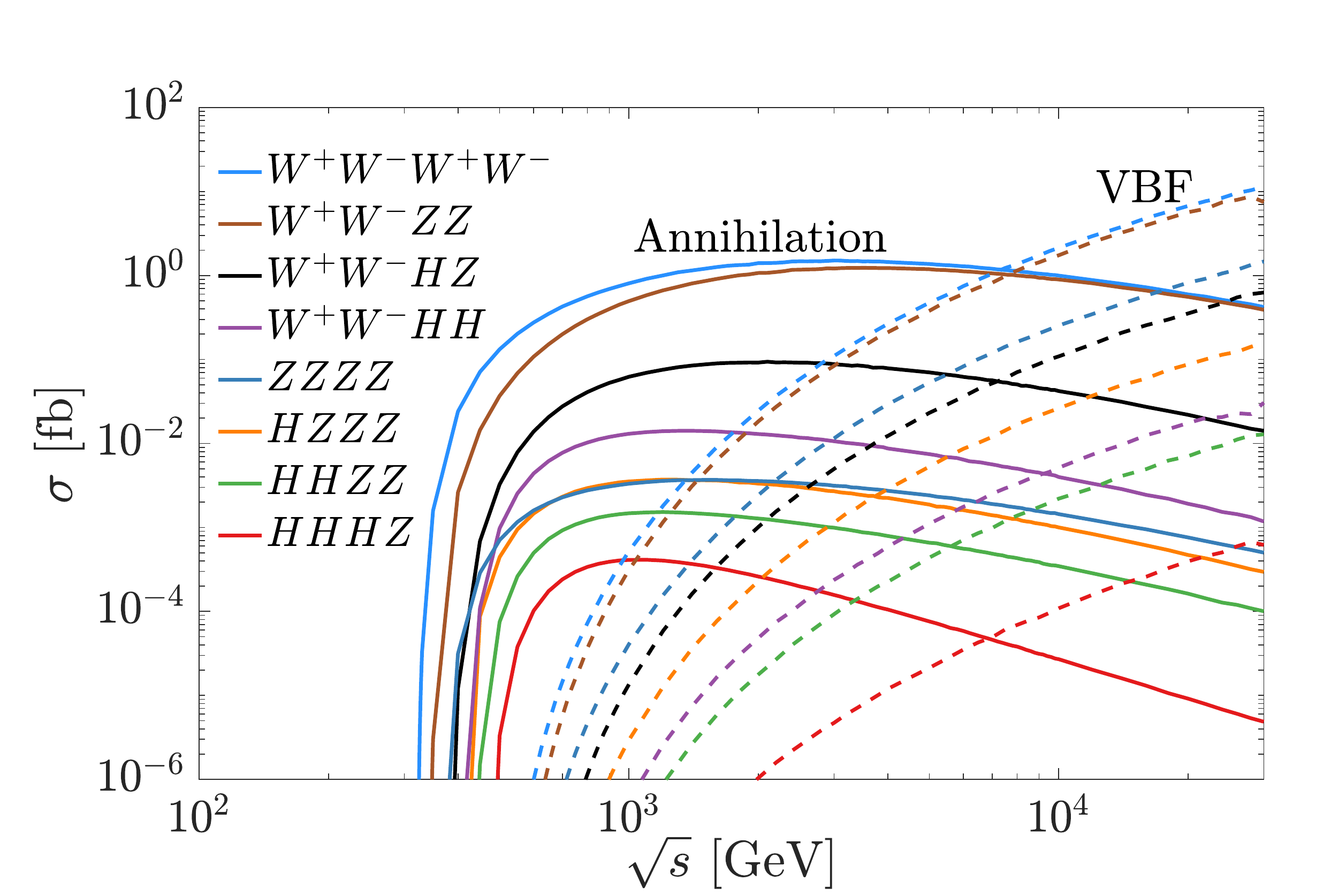}
\caption{Similar to Fig.~\ref{fig:2B}, the cross sections of four-boson production at a $\mm$ collider as a function of the c.m. energy $\sqrt{s}$, for SM $\kappa_\mu=1$ only.} 
\label{fig:4A}
\end{figure}

\begin{figure}
  \centering
  \includegraphics[width=0.8\textwidth]{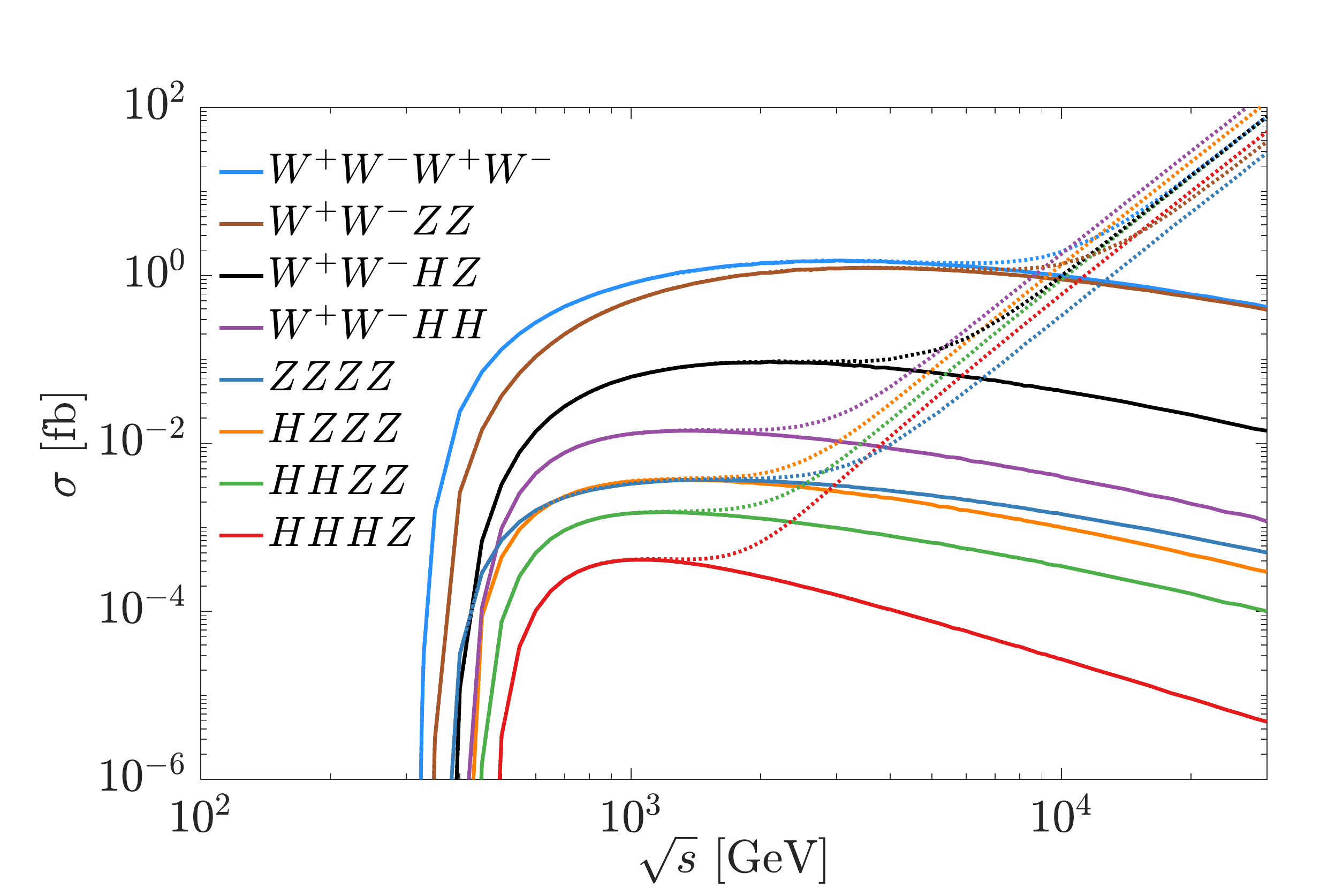}
  \includegraphics[width=0.8\textwidth]{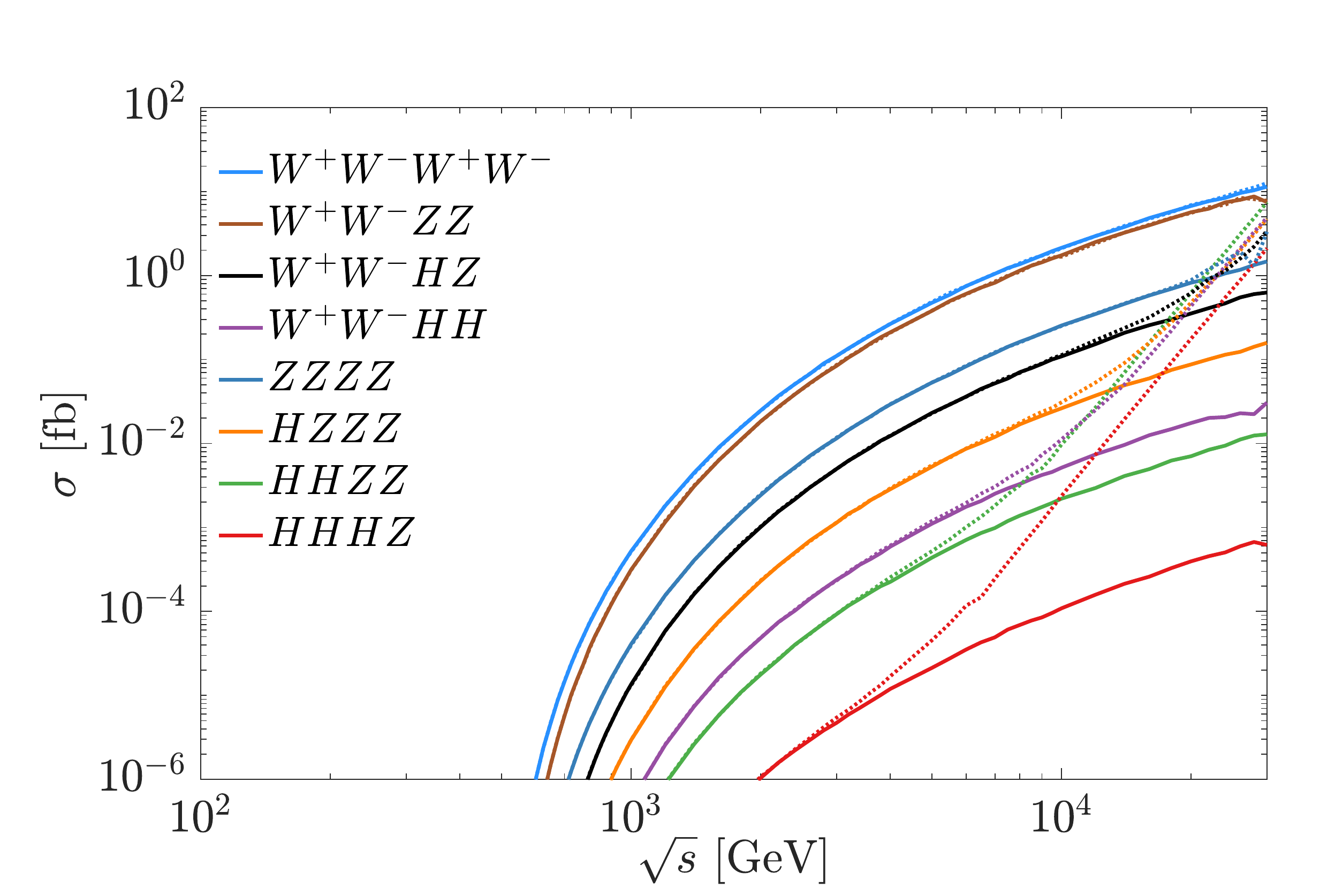}
  \caption{The cross sections of four-boson production at a $\mm$ collider via (a) annihilation $\mm \to 4B$ and (b) the (charged)
    vector boson fusions (VBF), $\mm\to\nn X$ as functions of the c.m. energy $\sqrt{s}$. The solid and dotted lines are for
      the results with muon Yukawa coupling as $\kappa_\mu=1$
      and $\kappa_\mu=0~(2)$, respectively. 
    } 
  \label{fig:4B}
\end{figure}

\subsection{Multi-boson production}

To numerically determine the different multi-boson production cross sections and later on assess the sensitivity to the muon Yukawa coupling,
we parameterize the EFT contributions discussed in the last section with a model-independent coupling $\kmu$, \emph{e.g.}, Eq.~(\ref{eq:kmu_heft}) or (\ref{eq:kmu_smeft}), and implement it into the multi-purpose event generator {\sc Whizard~2.8.5}~\cite{Kilian:2007gr,Moretti:2001zz,Brass:2018xbv} using its plugin to external models~\cite{Christensen:2010wz}. This is building upon the EFT frameworks used for multi-boson production and vector-boson scattering at hadron~\cite{Alboteanu:2008my,Kilian:2014zja,Brass:2018hfw,Ballestrero:2018anz} and electron-positron colliders~\cite{Beyer:2006hx,Fleper:2016frz}, which we adapted here for the muon collider. The QED initial-state radiation (ISR), resummed to all orders in soft photons and up to third order in hard-collinear radiation, is equally applicable to the muon collider. Beam spectra for multi-TeV muon colliders are much less complicated than for electron-positron colliders and can be easily described with a Gaussian beam spread of 0.1\%. They are, however, not relevant at the level of this study. 

\begin{figure}
    \centering
    \includegraphics[width=.25\textwidth]{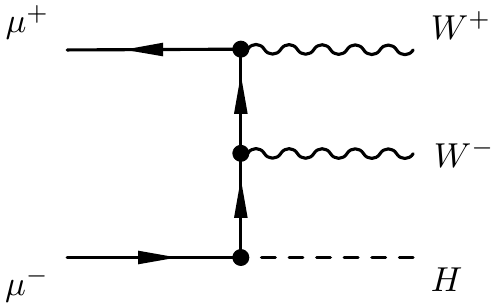}
    \quad
    \includegraphics[width=.25\textwidth]{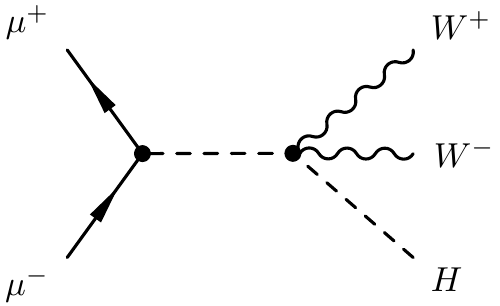}
    \quad
    \includegraphics[width=.25\textwidth]{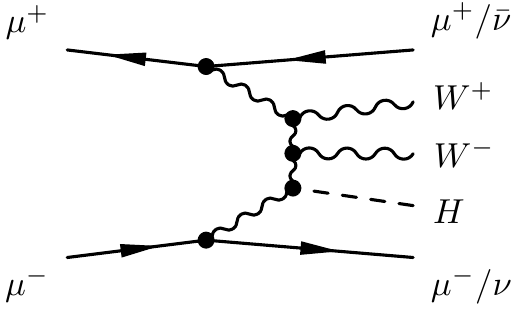}
    \caption{Representative diagrams for the signal annihilation process $\mu^+\mu^- \to W^+W^-H$}
    (left and middle), and for the VBF background process (right).
    \label{fig:mumuWWH}
\end{figure}

In Figs.~\ref{fig:2B}, \ref{fig:3B} and \ref{fig:4A}, we first present the Standard Model (with $m_\mu=y_{\mu}v/\sqrt{2}$) cross sections for the production of two, three and four bosons, respectively, including the Higgs and the EW 
gauge bosons. The cross sections -- in each case decreasing in size --  are 
for two-boson production,
\begin{equation}\label{eq:2B}
WW,~ZZ,~ZH,~HH
\end{equation}
for three-boson production,
\begin{equation}\label{eq:3B}
WWZ,~WWH, ~ZZZ, ~ZZH, ~ZHH,~HHH
\end{equation} 
and for four-boson production, 
\begin{equation}\label{eq:4B}
WWWW, ~WWZZ, ~WWHZ, ~WWHH, ~ZZZZ,~HZZZ, ~HHZZ,~HHHZ
\end{equation}
respectively. The single Higgs ($H$) production is also illustrated in Fig.~\ref{fig:2B}, which are obtained through $\mm\to H$ recoiled by ISR. We present two classes of production mechanisms, namely, the direct $\mm$ annihilation and the vector boson fusion (VBF) resulting from the initial-state radiation off the muon beams.\footnote{If no specific indication, we only include the charged vector boson ($W^{\pm}$) in VBF, \emph{i.e.}, $W^+W^-\to X$. The $Z$ boson fusion, $ZZ\to X$, is sub-leading due to its smaller vector coupling to leptons, with the example of $ZHH$ production demonstrated in Table \ref{tab:cutflow}. The final states involving charged particles, \emph{e.g.}, $W^+W^-H$, can be produced through photon or photon-$Z$ fusion as well, which are mostly collinear to the initial beams. This background is largely excluded when a reasonable angular cut (\emph{e.g.}, $10\degree<\theta<170\degree$) is imposed, also illustrated in Table  \ref{tab:cutflow}.} Representative  Feynman diagrams for these production mechanisms are shown in Fig.~\ref{fig:mumuWWH} for the $W^+W^-H$ final state. Near the threshold, the annihilation cross sections dominate. With the increase of collision energy, they are suppressed by $1/s$. The VBF mechanisms, on the other hand, increase with energy logarithmically \cite{Costantini:2020stv,Han:2020uid} and eventually take over above a few TeV. The $\mm$ annihilation to multiple Higgs bosons is induced by the Yukawa and possible Higgs self interactions, while no gauge couplings. The corresponding cross sections are highly suppressed compared with the channels involving gauge boson(s), with examples of $HH$ and $HHH$ demonstrated in Fig.~\ref{fig:2B} and \ref{fig:3B}.
Therefore, there is no need to include four-Higgs production in Eq.~(\ref{eq:4B}) or Fig.~\ref{fig:4B}, and the corresponding phenomenological study of the pure Higgs production is largely left for the future.

In the presence of anomalous couplings, the characteristic high-energy behavior shown in these figures is modified, 
as we discussed above in Sec.~\ref{sec:setup}. At asymptotically high energy, for each final state the new-physics contribution dominates
over the SM and exhibits a simple and uniform power law as shown in Figs.~\ref{fig:2B}, \ref{fig:3B} and \ref{fig:4B} by the dotted curves,
which behave as straight lines in double-logarithmic plots.

In Sec.~\ref{sec:setup} we provided a description within the EFT framework, in which the muon Yukawa coupling can receive contributions from new physics beyond the SM. 
The breakdown of the final states in terms of individual channels follows precisely the ratios of cross-section differences in Tables~\ref{tab:ratios-3}
and~\ref{tab:ratios-4}, respectively, for the matched model.  Given real
data, measuring those ratios at various energy values will allow us to deduce
the underlying pattern.  In particular, the absence of pure multi-Higgs states is a special feature for the extreme scenario $d\to\infty$ which we used
for the plots in Fig.~\ref{fig:3B}~and~\ref{fig:4B}, i.e., there are no direct 
muon-Higgs couplings at any order.  In a more generic scenario, multi-Higgs 
states will appear with a sizable rate, and the observable ratios of 
vector-boson and Higgs final states are related to the operator structure in 
the SMEFT expansion.

We now discuss the phenomenology of a modified muon Yukawa coupling in more detail. In the effective approach discussed above, the muon Yukawa coupling gets a modification like Eq.~(\ref{eq:kmu_heft}) or (\ref{eq:kmu_smeft}). In such a way, $\kmu=1$ corresponds to the SM case. The deviation of $\kmu$ from 1 quantifies the new physics contribution, which serves as the signal in this work. 
In Figs.~\ref{fig:3B}-\ref{fig:4B}, we showed two such benchmark cross sections for $\kmu=0$ and 2 as dotted curves. They coincide with each other, which reflects a symmetry of the annihilation cross sections such that 
\begin{equation}\label{eq:xsec}
 \sigma|_{\kmu=1+\delta}=\sigma|_{\kmu=1-\delta},
\end{equation}
where $\delta$ is the deviation from the SM muon Yukawa prediction, with an exception for the pure Higgs production. 

With $\kmu=0\ (2)$ at a high energy, the annihilation cross sections of
the $ZZH$ and $ZHH$ channels merge in Fig.~\ref{fig:3B}(a), which is a result of the Goldstone
equivalence between the longitudinal $Z$ boson and the Higgs. A similar situation happens to the four-boson case at a higher collision energy in Fig.~\ref{fig:4B}(b).  When compared with the Standard Model annihilation, we find that the
 $\kmu=0\ (2)$ cross sections agree at low collision energies, but gradually diverge as the collision energy increases. At $\sqrt{s}=30$ TeV, the relative cross section deviation can be three orders of  magnitude for the $ZHH$ case, while it amounts to 20\% for $WWZ$ case. This big difference provides
 us a good opportunity to test the muon Yukawa coupling at a
 multi-TeV $\mm$ collider.   

As discussed above, and pointed out in~\cite{Han:2020uid,Costantini:2020stv}, the annihilation process, in our particular case here for three-boson production, is overcome at high energies by the
vector-boson fusion (VBF) production which becomes dominant at all high-energy (lepton) colliders. Here we show the
VBF cross sections as dashed lines in Fig.~\ref{fig:3B}, as well. They are
calculated with the fixed-order approach for fusion processes
$\mm\to\nn X$, where $X$ represents the desired final-state
particles. We have imposed a cut on the invisible neutrinos,
$M_{\nn}>150$ GeV~\cite{Boos:1997gw,Boos:1999kj}, to suppress the on-shell decay $Z\to\nn$. We see that at an energy as high as 30 TeV, the VBF cross sections are generally $2\sim3$ magnitudes larger than the annihilation processes for three-boson production. The relative size is even larger for the four-boson case. These channels will serve as backgrounds for the annihilation multi-boson productions when we measure the muon Yukawa coupling. 

\begin{figure}
  \centering	
  \includegraphics[width=0.48\textwidth]{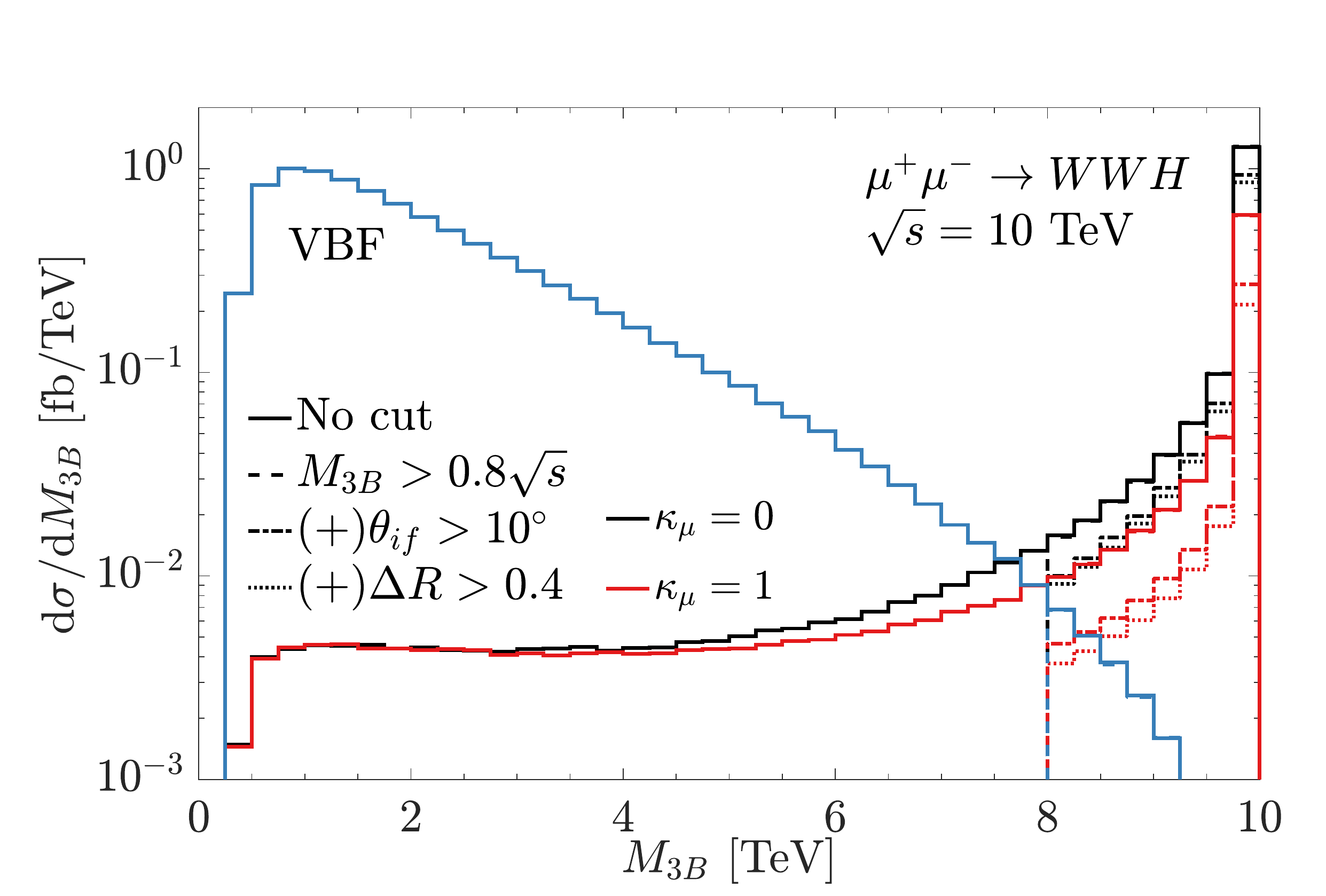}  
  \includegraphics[width=0.48\textwidth]{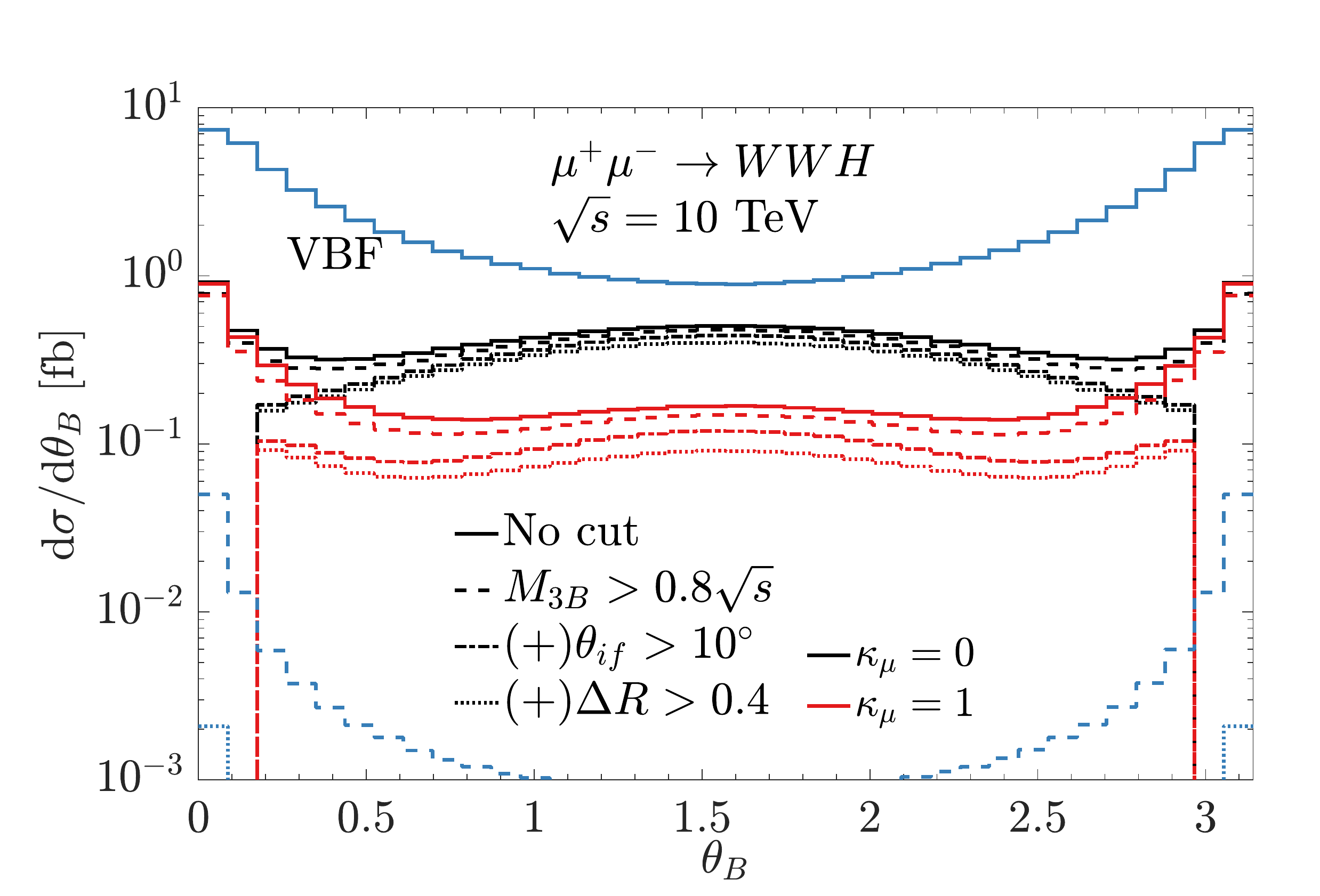}
  \includegraphics[width=0.48\textwidth]{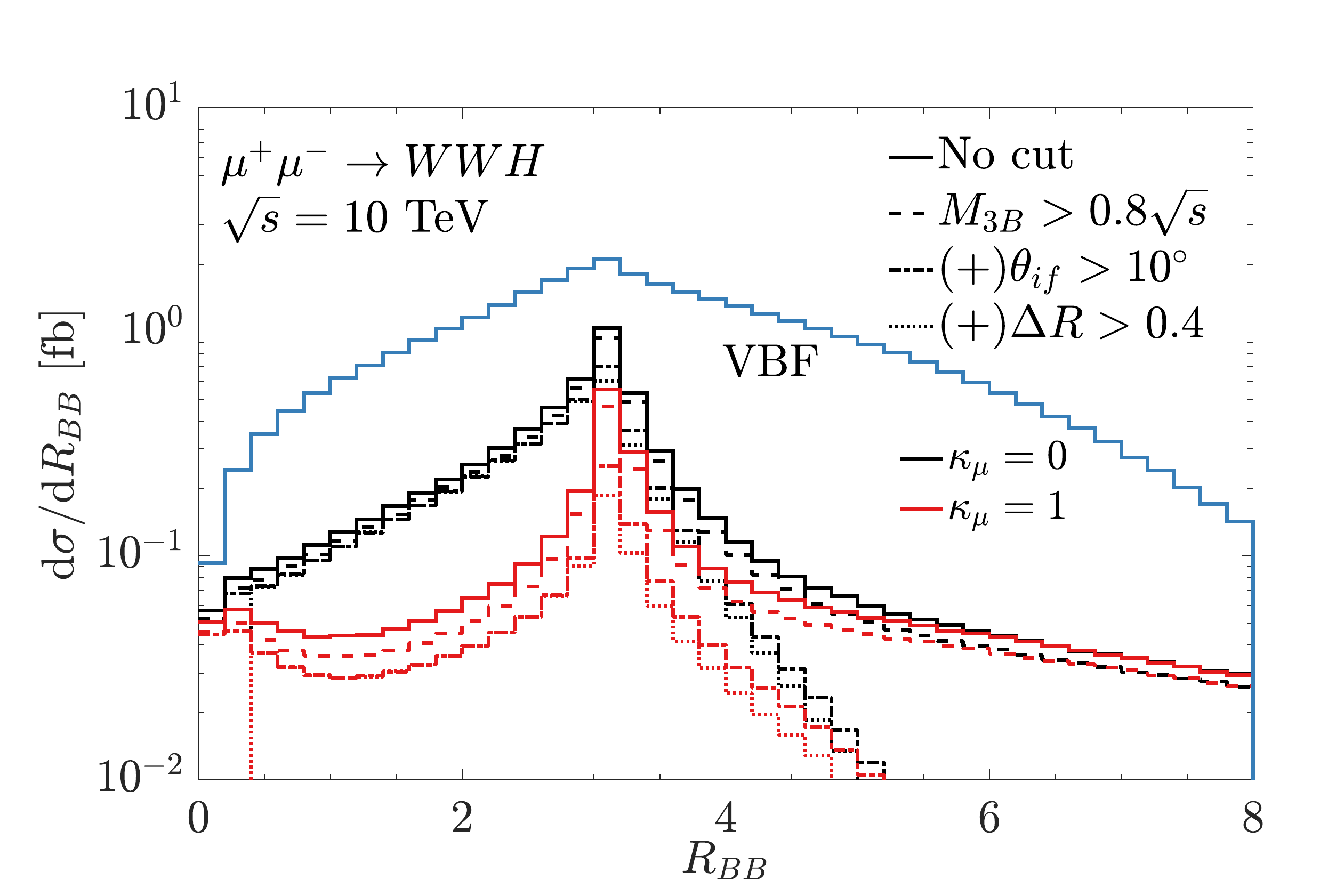}
  \caption{The kinematic distributions of the boson angle $\theta_B$, the diboson distance $R_{BB}$, and the triboson invariant mass $M_{3B}$ ($B=W,H$), respectively, in the $WWH$ production at a $\sqrt{s}=10$ TeV $\mm$ collider.} 
  \label{fig:distWWH}
\end{figure}

\begin{figure}
\centering	
\includegraphics[width=0.48\textwidth]{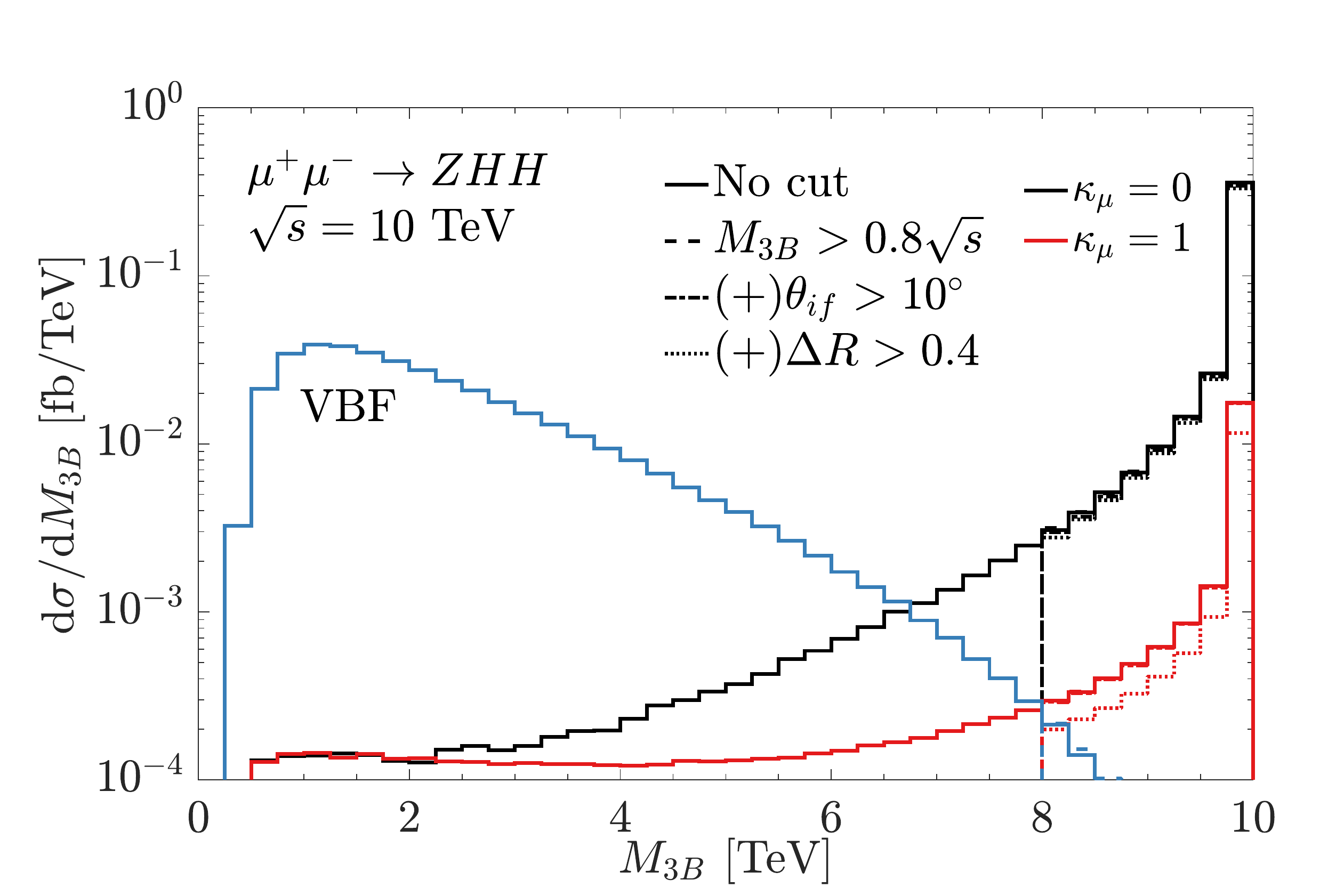}
\includegraphics[width=0.48\textwidth]{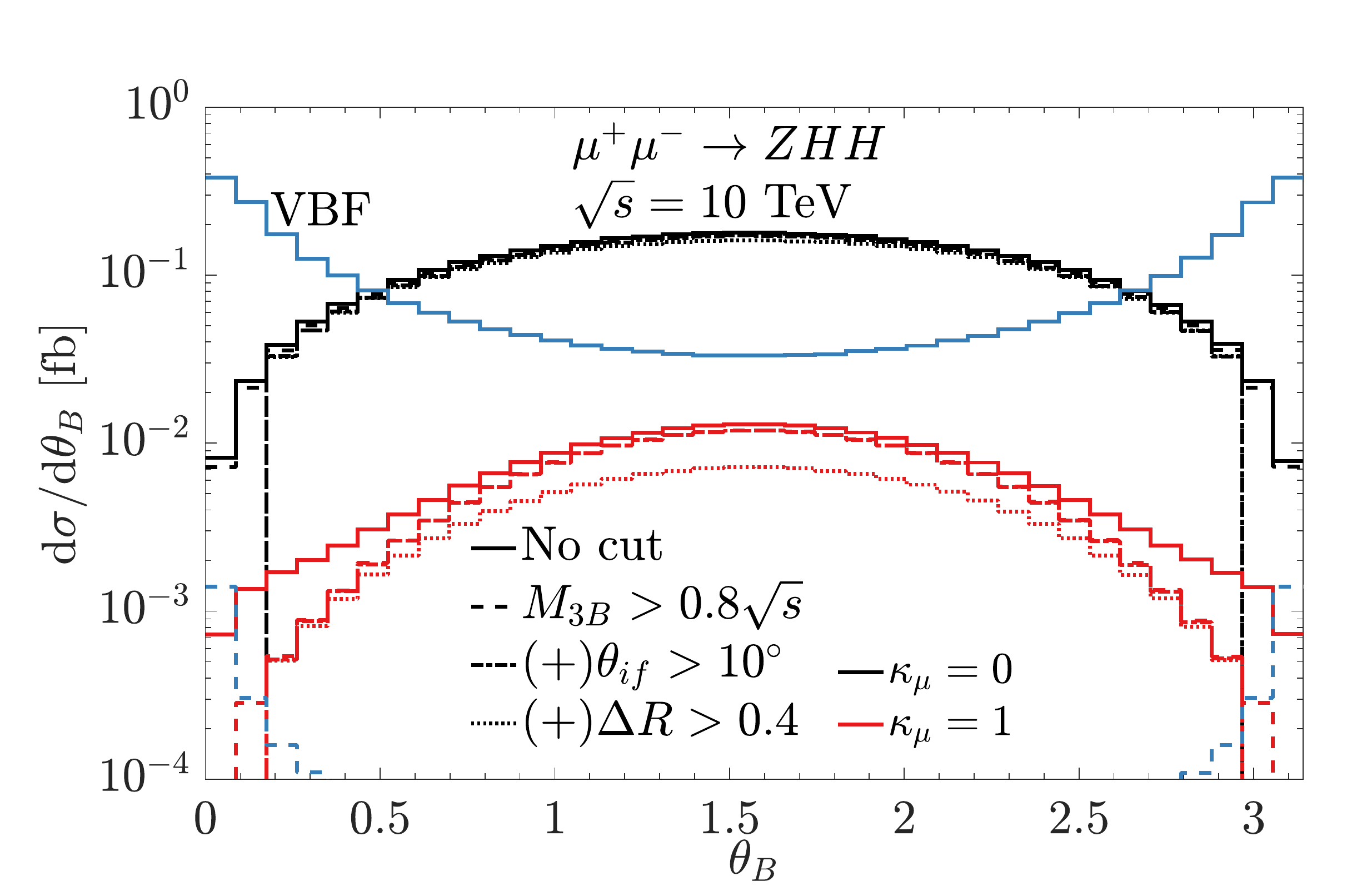}
\includegraphics[width=0.48\textwidth]{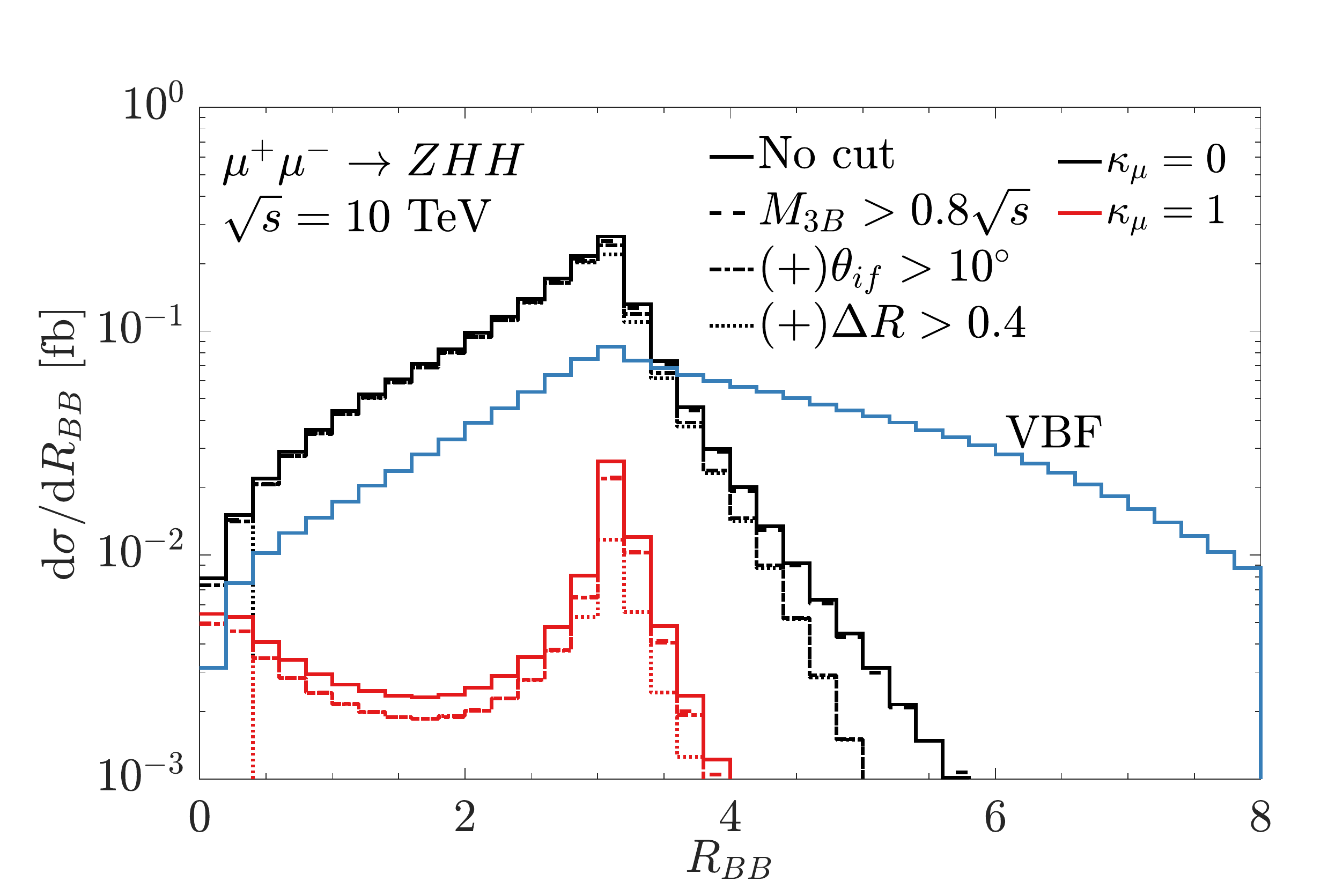}
\caption{The kinematic distributions for $\theta_B$, $R_{BB}$, and $M_{3B}$ as in Fig.~\ref{fig:distWWH}, but for $ZHH$ production at a $\sqrt{s}=10$ TeV $\mm$ collider.}
\label{fig:distZHH}
\end{figure}

\subsection{Kinematic distributions}
\label{sec:dist}
As we know, the kinematic distributions for the annihilation and VBF processes behave very differently. 
We take the $WWH$ and $ZHH$ production at a $\sqrt{s}=10$ TeV $\mm$ collider as benchmark examples\footnote{In triboson production, we choose $WWH$ as a demonstration example considering its large production rate, and $ZHH$ as another one for its relatively large deviation from the anomalous coupling. The $WWZ$ channel has an even larger cross section, while it suffers from a small relative deviation.} and show the distributions of boson angles $\theta_B\ (B=W,Z,H)$, the diboson separation distances $R_{BB}=\sqrt{(\Delta\eta)^2+(\Delta\phi)^2}$ in the rapidity-azimuthal angle plane, and triboson invariant masses $M_{3B}$, respectively, in
Fig.~\ref{fig:distWWH}~and~\ref{fig:distZHH}.
We see two main differences. First, the
invariant mass $M_{3B}$ for the annihilation process is sharply
peaked at the collision energy $\sqrt{s}$ seen in Fig.~\ref{fig:distWWH}(a) and \ref{fig:distZHH}(a), with a small spread due to the initial-state radiation (ISR). In contrast, in vector-boson fusion, the $M_{3B}$ is mainly peaked around the threshold. This feature enables us to efficiently separate these two processes and reduce the VBF background with an invariant mass cut. More specifically, with the $M_{3B}>0.8\sqrt{s}$
cut, the VBF background is reduced by three orders of magnitudes, with the absolute differential cross sections falling below the lower axis limits in Figs.~\ref{fig:distWWH}~and~\ref{fig:distZHH}.
In comparison, the signal, $\kmu=0~(2)$, almost remains the same size, with specific numbers listed in Tab.~\ref{tab:cutflow}. We also include the cut flow for the cross sections of SM annihilation to $WWH$ and $ZHH$ without including the ISR effect in Tab.~\ref{tab:cutflow}. We see the invariant mass cut does not impact at all in this case, because the $M_{3B}=\sqrt{s}$ is exact as a result of the momentum conservation. Another important observation is that the invariant mass cut $M_{3B}>0.8\sqrt{s}$ together with the ISR effect gives roughly the same cross sections without ISR, which justifies neglecting the ISR effect when necessary.

\begin{table}
\centering
	\begin{tabular}{c|c|c|c|c|c}
	\hline\hline
	Cut flow & $\kmu=1$ &  w/o ISR  & $\kmu=0~(2)$  & CVBF & NVBF  \\\hline \hline
	$\sigma$ [fb] &  \multicolumn{5}{c}{$WWH$} \\\hline
	No cut                            & 0.24 &  0.21  & 0.47 & 2.3                & 7.2  \\
	$M_{3B}>0.8\sqrt{s}$              & 0.20 &  0.21  & 0.42 & $5.5\cdot10^{-3}$  & $3.7\cdot10^{-2}$ \\
	$10\degree<\theta_{B}<170\degree$ & 0.092 & 0.096 & 0.30 & $2.5\cdot10^{-4}$  & $2.7\cdot10^{-4}$ \\ 
	$\Delta R_{BB}>0.4$               & 0.074 & 0.077 & 0.28 & $2.1\cdot10^{-4}$  & $2.4\cdot10^{-4}$  \\
	\hline 
	\# of events              & 740 & 770 & 2800 & 2.1 & 2.4  
	\\\hline
	$S/B$ & \multicolumn{5}{c}{2.8}   \\\hline\hline
	$\sigma$ [fb] & \multicolumn{5}{c}{$ZHH$}\\ \hline
	No cut                            & $6.9\cdot10^{-3}$ & $6.1\cdot10^{-3}$ & 0.119 & $9.6\cdot10^{-2}$  & $6.7\cdot10^{-4}$\\
	$M_{3B}>0.8\sqrt{s}$              & $5.9\cdot10^{-3}$ & $6.1\cdot10^{-3}$ & 0.115 & $1.5\cdot10^{-4}$  & $7.4\cdot10^{-6}$\\
	$10\degree<\theta_{B}<170\degree$ & $5.7\cdot10^{-3}$ & $6.0\cdot10^{-3}$ & 0.110 & $8.8\cdot10^{-6}$  & $7.5\cdot10^{-7}$\\
	$\Delta R_{BB}>0.4$               & $3.8\cdot10^{-3}$ & $4.0\cdot10^{-3}$ & 0.106 &$8.0\cdot10^{-6}$  & $5.6\cdot10^{-7}$ \\\hline
	\# of events & 38 & 40 & 1060 & -- & -- \\\hline
	$S/B$ & \multicolumn{5}{c}{27}   \\\hline\hline
\end{tabular}
 \caption{The cut-flow for the cross sections of $WWH$ and $ZHH$ production through annihilation (SM with $\kappa_\mu = 1$) with and without ISR, and the BSM signal models for $\kmu=0~(2)$ (i.e., $\Delta\kappa_\mu = \pm 1$). The last two columns are the SM backgrounds from charged (CVBF) and neutral vector boson fusion (NVBF), respectively. All cross sections are at a $\sqrt{s}=10$ TeV $\mm$ collider. The event numbers correspond to an integrated luminosity $\mathcal{L}=10~\textrm{ab}^{-1}$.
 The signal and background are defined in Eq.~(\ref{eq:SB}).
} 
\label{tab:cutflow}
\end{table}

Second, the final-state particles produced in the vector boson fusion are very forward, shown in Fig.~\ref{fig:distWWH}(b) and \ref{fig:distZHH}(b). In comparison, the annihilation-produced particles are much more central, especially for the events induced by a Yukawa interaction with $\kappa_\mu=0~(2)$. 
With an angular cut, such as $10\degree<\theta_{B}<170\degree$ based on the detector design \cite{Bartosik:2020xwr}, we are able to reduce
the VBF background by more than another factor of 10. The SM annihilation cross section will be suppressed by a factor of 2 for $WWH$, while the signal events with $\kmu=0~(2)$ are only reduced by 30\%. As for the case of the $ZHH$ processes, the impact of the angular cut is small both for the VBF background and for the annihilation process. 

Finally, in order to reasonably resolve the final states within
the detector, we need to require a basic separation among the reconstructed final-state bosons. The distributions of separation distance $R_{BB}$ in the $WWH$ and $ZHH$ production are shown in Fig.~\ref{fig:distWWH}(c) and \ref{fig:distZHH}(c). 
Besides the peak around $R_{BB}\sim\pi$ due to the back-to-back configuration, we obtain another minor peak around $R_{BB}\sim0$ for the SM annihilations, which reflects the collinear splitting behaviors, such as $W\to WH$ or $Z\to ZH$. With a reasonable separation cut $R_{BB}>0.4$, 
the SM annihilation to $ZHH$ is reduced by roughly 30\% due to the removal of radiation patterns with collinear splitting $Z\to ZH$. In comparison, both signal and backgrounds for $WWH$ production are only reduced slightly, with specific numbers presented in Table \ref{tab:cutflow}. 
In this case, the collinear splitting coincides with the forward beam region, which is already cut away by the angular acceptance.

\begin{figure}
\centering
\includegraphics[width=0.48\textwidth]{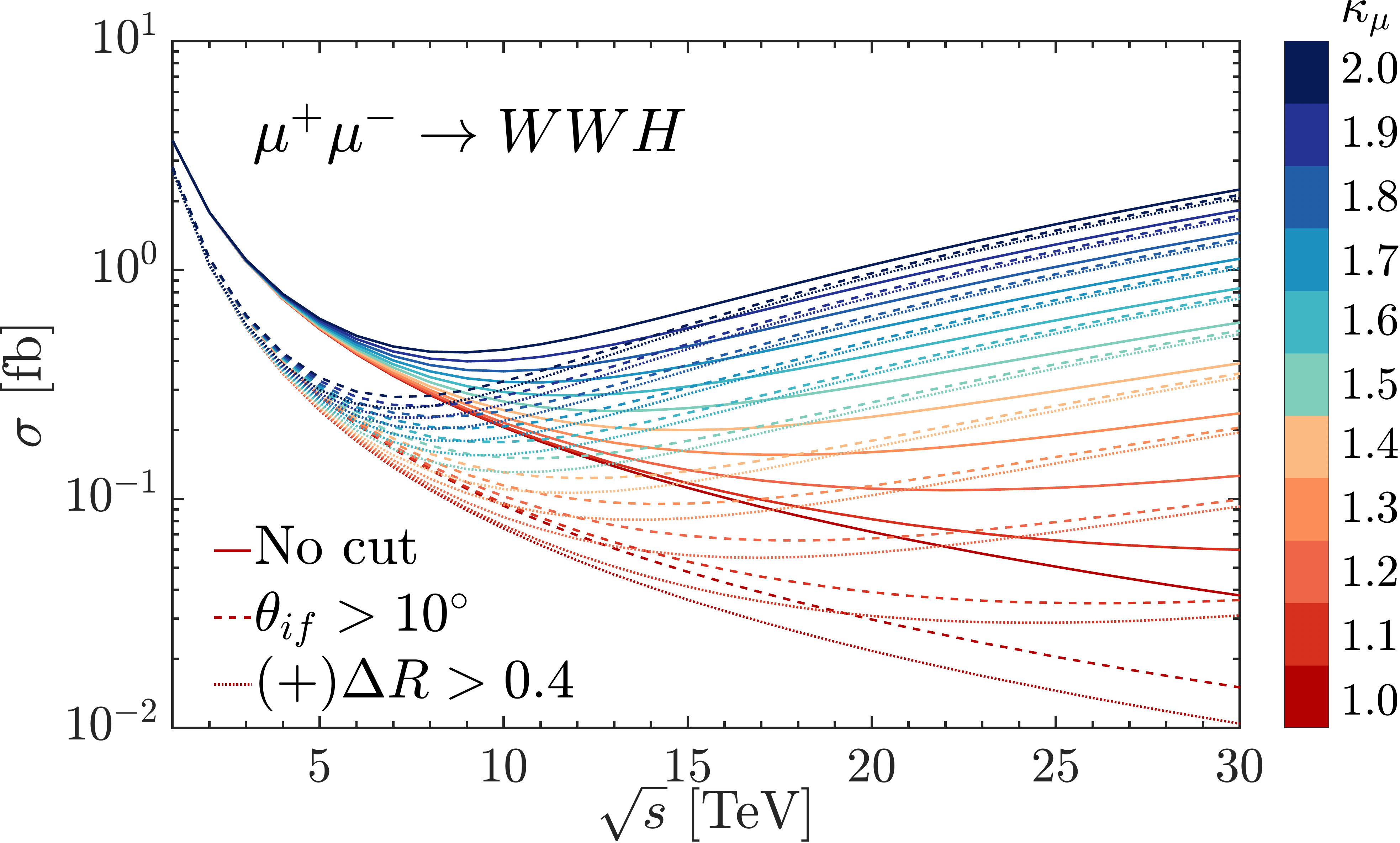}
\includegraphics[width=0.48\textwidth]{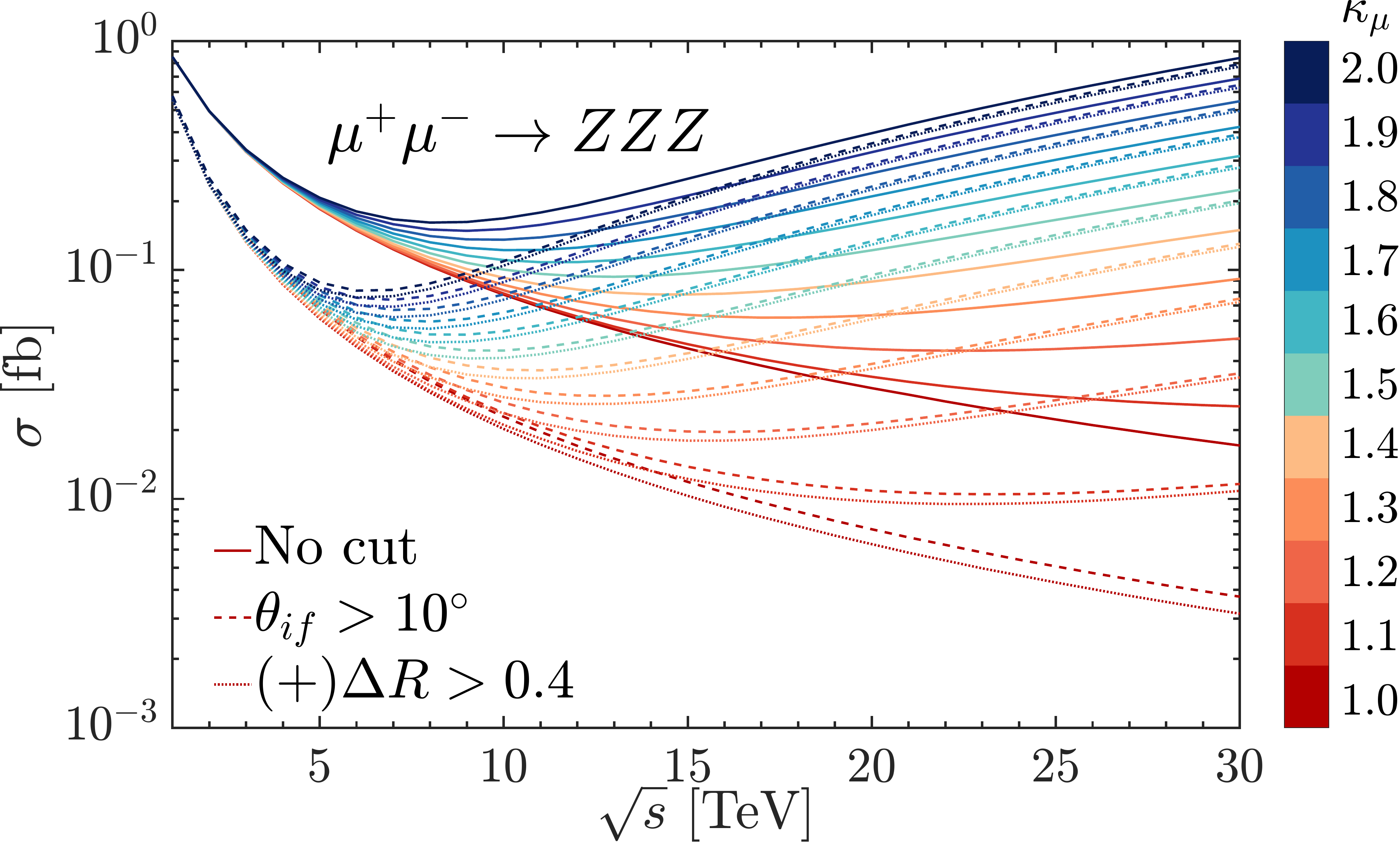}
\includegraphics[width=0.48\textwidth]{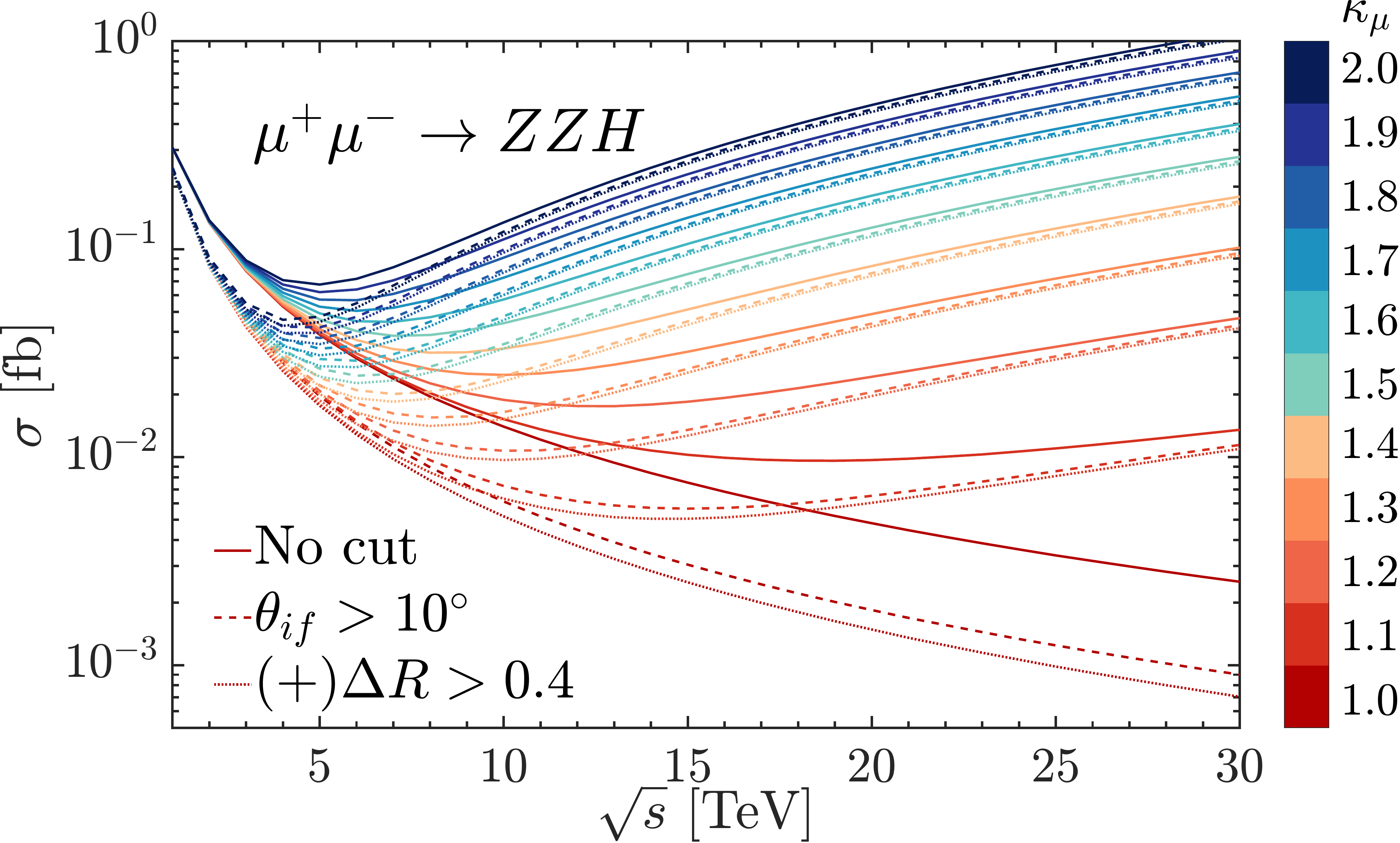}
\includegraphics[width=0.48\textwidth]{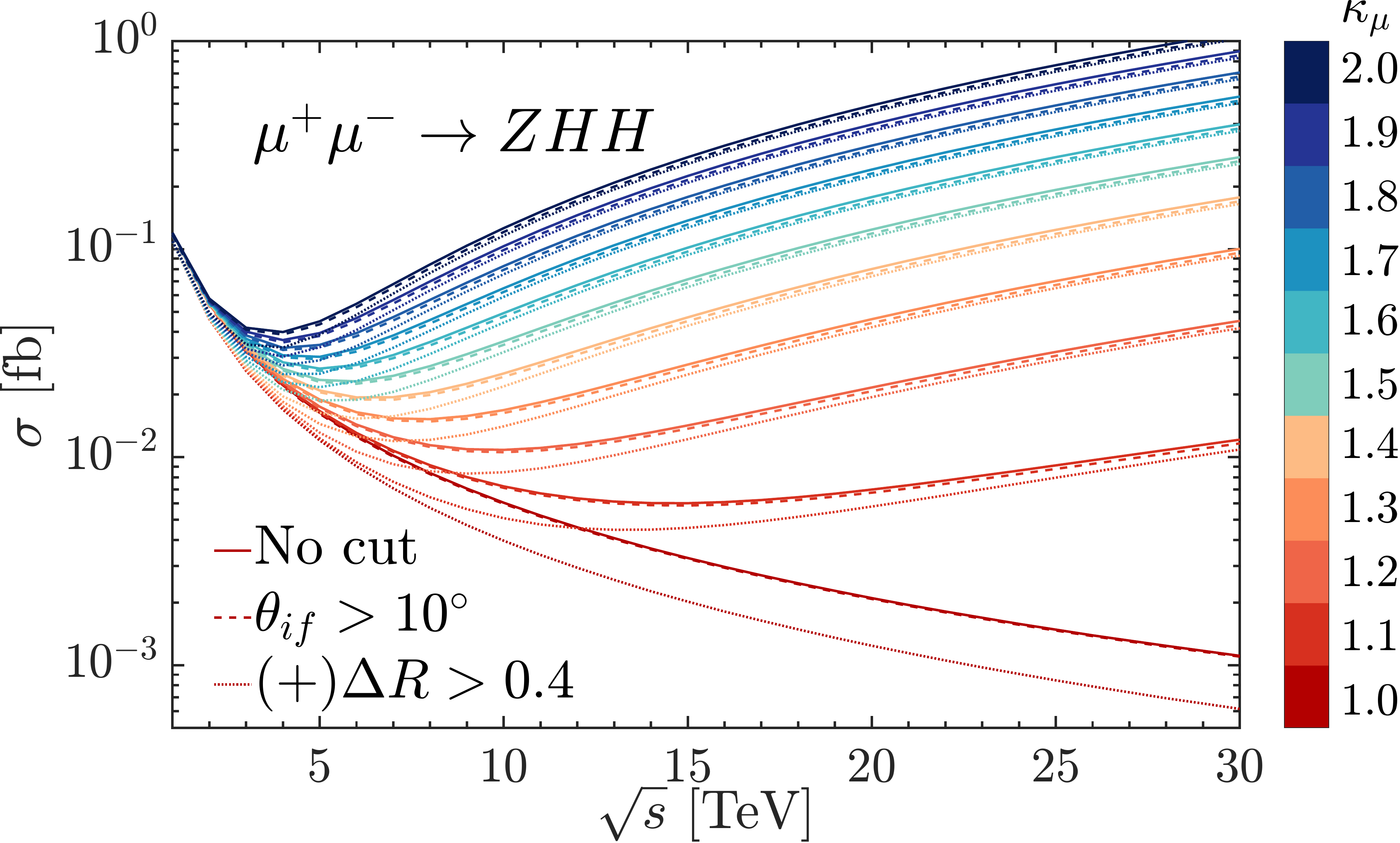}
\caption{The cross sections of annihilation without ISR for the three-boson production channels $\mm \to WWH, ZZZ, ZZH, ZHH$
  versus the $\mm$ c.m. energy $\sqrt{s}$ and the effective coupling $\kmu$. The lower two clusters of curves correspond the flow cut: $\theta_{if}>10\degree$ and the accumulated
  $\Delta R>0.4$.} 
\label{fig:scan}
\end{figure}

\subsection{Statistical sensitivity on the Muon Yukawa Coupling}

With the integrated luminosity in Eq.~(\ref{eq:lumi}), we obtain the event numbers for annihilation and VBF for $WWH$ and $ZHH$, listed in Table \ref{tab:cutflow}. We see a
big visible deviation from the SM backgrounds ($\kmu=1$) if we assume the muon Yukawa coupling varying within a range $\kmu=0 \ldots 1 \ldots 2$. We can obtain the signal and background events as
\begin{equation}\label{eq:SB}
S=N_{\kmu}-N_{\kmu=1}, ~B=N_{\kmu=1}+N_{\rm VBF},
\end{equation}
with a large signal-to-background ratio $S/B$ for $WWH$ and $ZHH$ shown in Table \ref{tab:cutflow}.
We can define the corresponding statistical sensitivity to the anomalous (non-SM) muon Yukawa coupling as 
\begin{equation}\label{eq:sensi}
 \mathcal{S}=\frac{S}{\sqrt{B}}.
\end{equation} 
We would like to emphasize that $\calS$ is always positive due to $N_{\kmu}\geq N_{\kmu=1}$, so we can define it without a modulus. We would expect a big
sensitivity under the assumption $\kmu=0~(2)$ for both $WWH$ and $ZHH$ channels, with the specific values even beyond the applicability of Gaussian approximation adopted in Eq.~(\ref{eq:sensi}).

We want to know how precisely we can measure the muon Yukawa coupling at a high-energy muon collider. For this task, we perform a scan of the annihilation cross
sections over the collision energy $\sqrt{s}$ and the effective
coupling $\kmu$, with results in the band of curves shown in Fig.~\ref{fig:scan}. We do not
include the $WWZ$ channel as the corresponding sensitivity is small resulting from the relatively small deviation shown in Fig.~\ref{fig:3B}. The ISR effect is safely discarded in this scan, thanks to the balance of the invariant mass cut, illustrated by the example of $WWH$ and $ZHH$ production in Table \ref{tab:cutflow}.
In Fig.~\ref{fig:scan}, we present three clusters of curves to illustrate the impact of the cut flow. The solid lines indicate the annihilation cross sections without any cuts. The lower clusters of dashed and dotted curves correspond to the angular cuts $10\degree<\theta_{B}<170\degree$ and the accumulated $\Delta R_{BB}>0.4$. We see that at large collision energy, the signal cross sections corresponding to $\kmu\neq1$ are not hampered by the kinematic cuts compared to the SM annihilation ones ($\kmu=1$). Especially at a large $\kmu$ deviation, such as $\kmu=0(2)$, the cross sections with and without selection cuts are more or less the same. The angular cut almost has no impact on the $ZHH$ channel, because both the $Z$ and $H$ boson are predominantly central in this channel, as mentioned above and shown in Fig.~\ref{fig:distZHH} (b). Instead, the separation distance cut reduces the SM annihilation rate by a factor of 30\%$\sim$40\%, due to the removal of collinear   splittings of $Z\to ZH$.

\begin{figure}\centering
\includegraphics[width=0.8\textwidth]{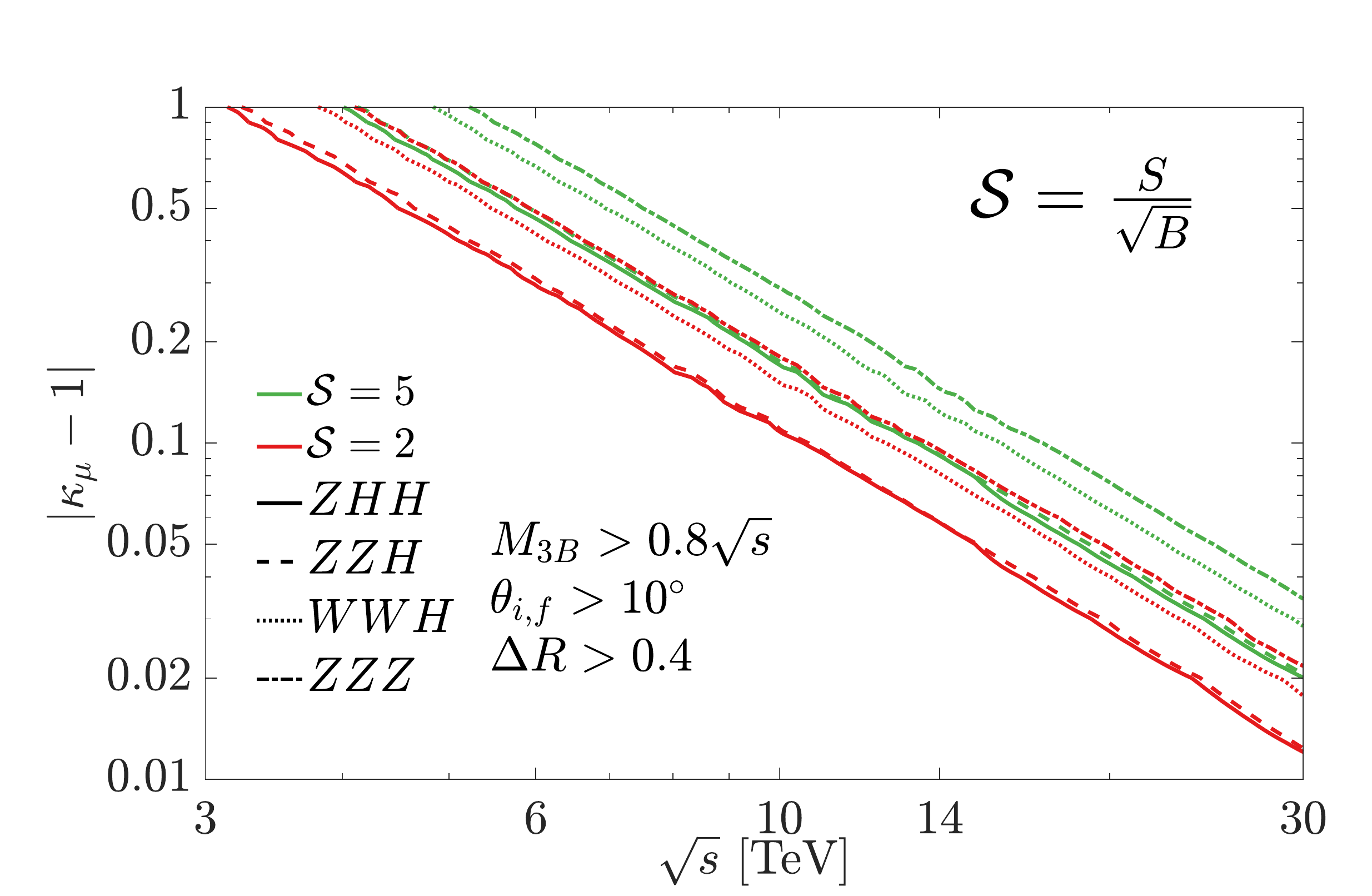}
\caption{The statistical sensitivity of a high-energy muon collider to the muon Yukawa coupling $\kmu$ from the measurements of three-boson production.} 
\label{fig:sensi}
\end{figure}

At this stage, we are able to obtain the sensitivity of a high-energy muon collider on the muon Yukawa coupling, by combining the cross sections with the
corresponding integrated luminosity. In Fig.~\ref{fig:sensi}, we show two type of contours, corresponding to $\calS=2$ and 5 respectively, with an integrated luminosity as given in Eq.~(\ref{eq:lumi}). We recall that the sensitivity respects a symmetry that $\calS|_{\kmu=1+\delta}=\calS|_{\kmu=1-\delta}$, due to the nature of the symmetric cross sections in Eq.~(\ref{eq:xsec}). The channels -- in decreasing size of sensitivity -- are $ZHH$, $ZZH$, $WWH$, and
$ZZZ$, respectively. At the low energy end, around 3 TeV, we are able to probe the muon Yukawa coupling about 100\% by means of the $ZHH$ channel, if we take the criterion $\calS=2$. At a 10 (30) TeV muon collider, we are able to test the muon Yukawa coupling to a precision of up to 10\% (1\%), mostly because of two factors: large signal-to-background ratios and large integrated luminosity. In addition, we see the
sensitivity of the $ZZH$ is very close to the $ZHH$ channel, as a result of the Goldstone equivalence theorem.   
Again, in the SMEFT formalism, the anticipated precision of $10\% - 1\%$ would translate to the sensitivity of the scale as 
$\Lambda \sim 30-100$ TeV.

So far in this paper, we have focused on the sensitivity to the muon Yukawa coupling from triboson production measurements at a high-energy muon collider. Similar analyses can be performed in the two- and four-boson channels. However, the sensitivities from the two-boson channels are expected to be weaker,  due to the relatively smaller sizes of the cross-section deviations from anomalous couplings, shown in Fig.~\ref{fig:2B}. Though in the four-boson channels, the signal-to-background ratios can be larger than that for the triboson channels, the production rates become significantly smaller compared to the three-boson channels. This elevates in our opinion the triple production to the ``golden channels" for this kind of measurement. Our event selection is based on imposing an invariant mass cut $M_{3B}>0.8\sqrt{s}$ in our analysis to enrich the annihilation channels. An opposite selection cut could likewise yield enriched samples of VBF processes; this is also expected to have some sensitivity on anomalous muon-Higgs couplings, based on the deviations shown in Fig.~\ref{fig:4B}(b). As a final remark, annihilation cross sections of (pure) multi-Higgs production do not respect the symmetry in Eq.~(\ref{eq:xsec}), which provides an opportunity to determine the sign of the deviation $\delta=\kappa_\mu-1$. Nevertheless, the production rate is so small that not even a single expected event survives the event selection, given the luminosity in Eq.~(\ref{eq:lumi}). The only chance lies in the single Higgs production with collision energy right on the Higgs mass threshold. We leave all these possibilities to future dedicated studies.

To summarize our results, a high-energy muon collider in the range of $10-30$ TeV, combining multi-TeV resolution power with the well-defined and clean leptonic environment, allows probing a tiny and elusive parameter of the SM like the muon Yukaww coupling to the single-digit percent level.

\section{Summary and Conclusions}
\label{sec:summary}

Motivated by the recent proposal for a multi-TeV muon collider, we explored the sensitivity of testing the muon-Higgs coupling at such a collider. Owing to the small muon-Yukawa coupling in the SM, any new physics contributions to the muon mass generation different from the SM Yukawa formalism would result in relatively large deviations from the SM prediction, and thus deserve special scrutiny at future collider experiments. We claim that a muon collider would be unique in carrying out such explorations. Our results are summarized as follows.

After presenting the  scale-dependence of the muon Yukawa coupling in the SM and in an extra-dimensional theory,
we discussed parameterizations for deviations of the muon-Yukawa coupling from its SM values within the frameworks of HEFT and SMEFT effective descriptions, and considered the implications on such anomalous couplings from perturbative unitarity bounds. As paradigm observables, we applied this EFT formalism to multi-boson production at a muon collider, particularly the production of two, three and four electroweak gauge bosons associated with a Higgs boson. Using the Goldstone boson equivalence theorem, we derived the scaling behavior of cross sections for processes with multiple bosons, containing deviations to the muon-Higgs coupling, normalized to specific reference cross sections for each multiplicity in Sec.~\ref{sec:ratios}. Our studies show that the sensitivity reach to such anomalous muon-Higgs couplings rises with the number of gauge bosons as the onset of the deviation from the SM is at lower energies. This is due to the fact that processes with higher multiplicities are involved in more insertions of the operators generating the deviations (and of higher operators) with high-energy enhancements and sizeable coupling coefficients. 

With the approach of a model independent effective coupling $\kappa_\mu$, we further performed detailed numerical analyses in  Sec.~\ref{sec:Pheno}, and found that two-boson production processes have less sensitivity to the muon-Yukawa coupling, while those for four-boson production have lower production rates. Therefore, to demonstrate the feasibility of such a study, we identified the optimal processes of triboson production $\mm\to W^+W^-H,ZHH$ as prime examples and showed how to isolate this from its most severe background, the same final state produced in vector-boson fusion. Typical observables are diboson correlations, either their invariant masses, their angular distributions or their $\Delta R$ distances. In this scenario, a muon collider with up to 30 TeV center-of-mass energy has a sensitivity to deviations of the muon-Yukawa coupling from its SM value of the order of 1\%$\sim$4\%. This can be interpreted in the SM as a measurement of the muon Yukawa coupling with this precision.
In the SMEFT formulation, if we assume an order-1 coupling, this precision would correspond to a probe to a new physics scale of  about $\Lambda \sim 30-100$ TeV.

There are many ways such an analysis can be improved, {\it  e.g.,}~by combining different channels, performing measurements at different energy stages of the machines, by combining final states with different multiplicities, by using multivariate analyses instead of simple cut-based analyses and by using polarization information on the final-state vector bosons. All of this is beyond the scope of this paper and is left for future investigations.

This paper highlights the tantamount possibilities to study one of the most elusive parameters within particle physics, the Higgs-muon coupling, and it also shows in more general context how effective field theories can be utilized to make the utmost use of a discovery facility like the muon collider.

%
%

\acknowledgments

We thank Fabio Maltoni, Daniel Schulte and Andrea Wulzer for useful discussions.  
This work was supported in part by the U.S.~Department of Energy under grant No.~DE-FG02-95ER40896, U.S.~National Science Foundation under Grant No.~PHY-1820760, and in part by the PITT PACC. 
JRR acknowledges the support by the Deutsche Forschungsgemeinschaft (DFG, German Research Association) under Germany's Excellence Strategy-EXC 2121 ``Quantum Universe"-39083330. WK and NK were supported in part by the Deutsche Forschungsgemeinschaft (DFG, German Research Foundation) under grant
396021762 – TRR 257.

%

\bibliographystyle{JHEP}
\bibliography{refs.bib}

\end{document}